\documentclass[a4paper,10pt]{article}

\usepackage{color}
\usepackage{amsthm, amsmath, amssymb, stmaryrd, dsfont}
\usepackage{physics}
\usepackage{xspace}
\usepackage{natbib}
\usepackage{subfigure}
\usepackage{authblk}
\usepackage{todonotes}
\usepackage{float} 
\usepackage{circuitikz}
\usetikzlibrary{decorations.pathmorphing}

\usepackage{tcolorbox}
\usepackage{tikz}
\newtcolorbox{hypothesis_box}[2][]{%
  colback=white, 
  colframe=black, 
  boxrule=0.3mm, 
  width=0.9\textwidth,
  left=3mm, 
  right=3mm, 
  top=2mm, 
  bottom=2mm, 
  center,
  arc=0mm,
  fonttitle=\bfseries,
  title=#2, 
  coltitle=black,
  colframe=gray!40,
  #1
}

\usepackage{hyperref}
\usepackage{comment}
\usepackage{multirow, multicol, caption, makecell, nicematrix, diagbox}
\usepackage[graphicx]{realboxes}
\usepackage{subcaption}
\hypersetup{
colorlinks=true,
breaklinks=true,
urlcolor= blue,
linkcolor= black, 
citecolor=black} 

\newcommand{\mrel}{m_\textup{rel}}
\newcommand{\mabs}{m_\textup{abs}}

\numberwithin{theorem}{section}

\title{A spatio-temporal graph-based model for team sports analysis}
\date{}
\author[1,2*]{Camille Grange}
\author[3]{Quentin Bourgeais}
\author[1]{Rodolphe Charrier}
\author[4]{Géraldine Del Mondo}
\author[1]{Antoine Dutot}
\author[1]{Eric Sanlaville}
\author[3,5]{Ludovic Seifert}

\affil[1]{\small{Université Le Havre Normandie, Univ Rouen Normandie, INSA Rouen Normandie, Normandie Univ, LITIS UR 4108, F-76600 Le Havre, France, email: \{rodolphe.charrier, antoine.dutot, eric.sanlaville\}@univ-lehavre.fr}}
\affil[2]{\small{Université Sorbonne Paris Nord, LIPN, CNRS, UMR 7030, F-93430, Villetaneuse, France, $^*$Corresponding author email: grange@lipn.univ-paris13.fr}}
\affil[3]{\small{Université de Rouen Normandie, CETAPS UR3832, Rouen, France, email: \{quentin.bourgeais, ludovic.seifert\}@univ-rouen.fr}}
\affil[4]{\small{INSA Rouen Normandie, Univ Rouen Normandie, Université Le Havre Normandie, Normandie Univ, LITIS UR 4108, F-76000 Rouen, France, email: geraldine.del\_mondo@insa-rouen.fr}}
\affil[5]{\small{Institut Universitaire de France, Paris, France}}

\begin{document}
\maketitle


\sloppy

\begin{abstract}
Team sports represent complex phenomena characterized by both spatial and temporal dimensions, making their analysis inherently challenging. In this study, we examine team sports as complex systems, specifically focusing on the tactical aspects influenced by external constraints. To this end, we introduce a new generic graph-based model to analyze these phenomena. Specifically, we model a team sport's attacking play as a directed path containing absolute and relative ball carrier-centered spatial information, temporal information, and semantic information. We apply our model to union rugby, aiming to validate two hypotheses regarding the impact of the pedagogy provided by the coach on the one hand, and the effect of the initial positioning of the defensive team on the other hand. Preliminary results from data collected on six-player rugby from several French clubs indicate notable effects of these constraints. The model is intended to be applied to other team sports and to validate additional hypotheses related to team coordination patterns, including upcoming applications in basketball.\\

\noindent\textbf{Keywords:} Modelization, graphs, team sports, spatial dimension, temporal dimension, experimental data analysis

\end{abstract}

\section{Introduction}

Nowadays, the study of complex systems is gaining popularity in the computer scientific community in order to analyze dynamic interactions between entities. A large part of such systems involve spatial and temporal considerations, namely spatial and temporal relationships between the different entities, also referred to as spatio-temporal (ST) phenomena. Note that merely defining spatial and temporal notions and relations is done in numerous different ways (e.g. absolute or relational concept of time, qualitative or quantitative temporal relationships, objects or fields concepts of basic unit of space, etc.). Interested readers can refer to~\cite{Oberoi2019} for a detailed overview of these notions. 

A large part of ST phenomena is currently modeled with graphs, encompassing a wide variety of domains and motivations such as modeling, visualizing, storing, extracting information, and simulating real-world phenomena. More precisely, ST graph-based models are used in fields such as Geographic Information Systems (GIS)~\citep{Spery2001, Sriti2005, DelMondo2013, DelMondo2010}, urban traffic~\citep{Oberoi2019, Costes2015, Ding2004} and knowledge representation~\citep{Aydin2016, VonLandesberger2015, Jiang2009, Lee2005}, facilitating for instance data visualization and information storage. They also play a significant role in extracting information through learning and object recognition, frequent patterns identification, anomalies detection and clustering. These applications cover multiple domains, among others computer vision and robotics~\citep{Sridhar2010, Toumpa2023, Sridhar2008, Jain2016}, GIS~\citep{Wu2021}, mobility~\citep{Maduako2019, Douillard2010, Jin2023} and criminology~\citep{Davies2015, Sanabria2022}. Additionally, ST graph-based models are used for simulation purposes in epidemic studies~\citep{Koher2019} to analyze complex situations and thus their ability to support informed decision-making.

In the above-mentioned models, vertices and edges have various definitions. On the one hand, vertices can represent objects, with or without spatial print, events, spatial and temporal relations, etc. On the other hand, edges (or arcs) can symbolize temporal relations, e.g. of filiation, spatial relations, spatio-temporal relations and thematic relations. Eventually, vertices and edges can be augmented with characteristics such as quantitative attributes or semantic. This rich variety of possibilities allows for the definition of numerous models tailored to the specific use case being addressed.\\

In this paper, we propose a new ST graph-based model for team sports in order to study the effect of some external parameters on team coordination patterns. 
Team sport can be described as the confrontation of two teams of collaborating players engaged in a competitive relationship, making it relevant to consider them as two interacting organized systems in a continuous opposition governed by spatio-temporal stakes~\citep{grehaigne1997}. The analysis of sport teams as complex systems has progressively gained attention in sport sciences~\citep{balague2013,davids2014book}, and graph theory has particularly emerged as a valuable tool in this context to model \emph{coordination patterns} within teams~\citep{passos2011}. This far, the main approach consists in modeling passing network~\citep{Buldu2018}, considering players as vertices linked by their successful passes. 

This systemic approach becomes particularly relevant when considering sport teams as \emph{complex adaptive systems}, whose behavior emerges from the ongoing interactions between players and their environment~\citep{davids2014}. More precisely, the characteristics of the task, of the environment and of the organism interact to form a set of \emph{constraints}, acting as boundaries that limit how behavior can emerge~\citep{newell1986}. These constraints evolve and create an unpredictable environment to which individuals or teams must continuously adapt~\citep{seifert2013}. In that perspective, coordination patterns are seen as an adaptive behavior of the team in response to a set of surrounding constraints, making the understanding of how teams adapt to such constraints a matter of interest. For example, the analysis of ball passing interactions among players through the use of temporal graph models revealed that basketball teams adapt their coordination patterns to the current score of the game, and that a team using more various coordination patterns is likely to score more points during the game~\citep{bourgeais2024temporal}. 

In order to better capture the real nature of the game, several ST graph-based models have been proposed in the literature: models that consider interactions between players occurring within specific zones of the field and at given moments of the game~\citep{cotta2013}; models that characterize passing sequences while incorporating their spatial location~\citep{wang2015}; models based on subgraphs within passing sequences, which also integrate spatial information about the ball-passing events~\citep{meza2017,barbosa2022}. Recently, \cite{oberoi2022} proposed a model providing a more complete description of the phenomenon: they define a ST graph-based model in which vertices represent individual players, field zones and the ball, and where edges represent the diverse relationships among those vertices; and they account for the temporal sequence of graphs.
Overall, in previous literature the spatial dimension has been integrated in various ways, including labeling vertices according to their tactical positions (e.g. attacker, defender), discretizing the field into zones, or directly using spatial 2-dimensional Euclidean coordinates of events. The temporal dimension, on the other hand, has been approached at different scales – ranging from game-level dynamics to passing sequence dynamics – and two main modeling strategies can be identified: one based on time discretization (i.e. time windows) and the other based on changes in graph topology (i.e. modifications in vertices, links, or labels). 

In the present work, we define a model based on a spatial discretization and on temporal changes in network topology, with the specificity of a spatial information ball carrier-centered. This is motivated by the study of the emergence of coordination patterns under various experimental constraints.\\

The structure of this paper is as follows. 
In Section~\ref{sec:spatiotemporalinfo}, we describe the nature of spatio-temporal information represented in the proposed model. In Section~\ref{sec:genericmodel}, we define a generic spatio-temporal graph-based model for team sports analysis. In Section~\ref{sec:appli_rugby}, we apply it for rugby and in Section~\ref{sec:experiments}, we state hypotheses on rugby union team coordination patterns and study their validity with the model. Eventually, we discuss the model and perspectives in Section~\ref{sec:discussion}.

\section{Spatio-temporal information}
\label{sec:spatiotemporalinfo}
In this section, we outline the representation of spatio-temporal information for the model described in Section~\ref{sec:genericmodel}. The chosen methods for representing space and time are based on the specific primordial information we aim to extract from the observed spatio-temporal phenomenon, particularly in the context of team sports. These choices were made in collaboration with team sports experts to ensure relevance and accuracy. One of the challenges is to find a balance between capturing sufficient characteristics of the phenomenon while avoiding excessive detail to maintain parsimony. In other words, we aim to extract relevant information to study team coordination patterns from the raw spatio-temporal data, such as the positions of each player on the field, without overwhelming the model with unnecessary complexity.

The model contains three types of information. Spatial information, which characterizes the spatial position of the players on the field; thematic information, which expresses remarkable events (e.g. a pass); and temporal information, which reflects the evolution of the game and is linked to spatial positions.

Henceforth, we note $n\in\mathbb{N}^*$ the number of players of the team. We use the notation $[n]:= \{1,\ldots,n\}$.

\subsection{Spatial information}
Let us begin with the spatial information we consider in the model. 
This information is composed of two indicators: the absolute and the relative spatial position. In order to define them, we call a \emph{zone} a connected bounded part of the Euclidean space of two dimensions. 
Let $A$ be a zone and $(B_i)_{i\in \mathcal{I}}$ be a finite set of zones. We recall that $(B_i)_{i\in \mathcal{I}}$ is a partition of $A$ if 
\begin{equation*}
    \bigcup_{i\in \mathcal{I}} B_i = A~~\text{ and }~~\bigcap_{i\in \mathcal{I}} B_i = \emptyset\,.
\end{equation*}
Henceforth, we consider the field as a zone called $F$.

First, we define the \emph{absolute spatial position}.
We partition $F$ into $\mabs$ zones $(A_i)_{i\in [\mabs]}$. This partition does not change over time and represents an absolute reference over the field. We call each $A_i$ an \emph{absolute} zone. We note $$\mathcal{A} = \{A_1,\ldots,A_{\mabs}\}$$ the set of absolute zones.
For a given instant time, we define the absolute spatial position of the game as $A\in\mathcal{A}$. The absolute position indicates in which absolute zone is the ball carrier at this instant time.

Second, we define the \emph{relative spatial position}.
For a given instant time $t\in\mathbb{R}^+$, we partition the zone $F\setminus\{pos(t)\}$, where $pos(t)\in F$ is the position of the ball carrier at time $t$, into $\mrel$ zones $(R_j(t))_{j\in [\mrel]}$. Thus, this partition evolves over time according to the position of the ball carrier. We call each $R_j(t)$ a \emph{relative} zone. Notice that the function that splits the field into $\mrel$ zones is the same for any position $pos(t)$, but the resulting splitting will differ according to the ball carrier position.
Thus, for a given instant time, we define the relative spatial position as a tuple $(N_1, N_2, \ldots, N_{\mrel})$ such that $\sum_{j=1}^{\mrel} N_j = n-1$ and $N_j\in\mathbb{N}, \forall j\in [\mrel]$. The relative position indicates the number of players (other than the ball carrier) in each relative zone. Precisely, $N_j$ is the number of players in zone $R_j(t)$. 
 We note $$\mathcal{R} = \{(N_1,\ldots, N_{\mrel}) : \sum_{j=1}^{\mrel} N_j = n-1, N_j\in \mathbb{N}\}$$ the set of all possible relative positions. 
 
Notice that, in the proposed model, we define the relative spatial position as being related only to the other players of the same team because we aim to study team coordination patterns, which are internal tactical effects. However, one can readily extends the definition to integrate the players of the opposite team (possibly with a different partitioning of the field), if studying the team's reaction to the opposite team for instance.
 
\paragraph{Spatial relations.} 
According to the definition of spatial information, two spatial positions can be easily compared, or linked, by examining whether they share the same absolute position and/or the same relative position. Specifically, we define the \emph{spatial relation} between two spatial positions as follows.
Given two spatial positions $(r_1, a_1), (r_2, a_2)\in \mathcal{R}\times \mathcal{A}$, $(r_1, a_1)\neq(r_2, a_2)$, their spatial relation is defined by a label $\rho_\text{sp} \subseteq \mathcal{R}_\text{sp}$, where

$$\mathcal{R}_\text{sp} = \{\text{rel}, \text{abs}\}\,.$$
Specifically, $\rho_\text{sp}$ indicates the nature of the spatial relation:

\begin{equation*}
    \rho_\text{sp} = \begin{cases} 
    \{\text{rel}\},~~\text{if}~~ r_1 \neq r_2 ~~\text{and}~~ a_1 = a_2~~ ~~\text{(relative position change only)}\\
    \{\text{abs}\},~~\text{if}~~ a_1 \neq a_2 ~~\text{and}~~ r_1 = r_2 ~~ ~~\text{(absolute position change only)}\\
    \{\text{rel, abs}\},~~\text{if}~~ r_1 \neq r_2 ~~\text{and}~~ a_1 \neq a_2~~ ~~\text{(both relative and absolute position changes)}
    \end{cases}
\end{equation*}

\subsection{Thematic information}
In addition to spatial information, we also consider \emph{thematic information}, which refers to remarkable events that we want to track and observe during team sports. Specifically, thematic information consists of elements from a set $\{th_1,\ldots,th_k\}$, for $k\in\mathbb{N}$. This set represents all the thematic events of interest for a particular sport, making it unique to each sport. 

One common thematic event across all team sports is a pass. This event can be enhanced with various semantic details, as illustrated in the rugby application discussed in Section~\ref{sec:appli_rugby}. This is important to note that a pass has a non-zero duration and typically results in a change of spatial position. However, we can also consider instantaneous events, such as a non-verbal communication through eye contact between two players to coordinate a play. Moreover, notice that a thematic event, whether instantaneous or not, may not result in any spatial change.

\paragraph{Thematic relations.} 
Given a thematic event, starting in spatial position $(r_1, a_1)\in \mathcal{R}\times \mathcal{A}$ and ending in spatial position $(r_2, a_2)\in \mathcal{R}\times \mathcal{A}$ (not necessarily different from $(r_1, a_1)$), the thematic relation between these two positions is defined by a label $\rho_\text{th} \subseteq  \mathcal{R}_\text{th}$, where 
$$\mathcal{R}_\text{th} = \{th_1,\ldots,th_k\}$$
is assimilated to the set of thematic events. 

\subsection{Temporal information}
We consider \emph{temporal information} as information exclusively related to a spatial position, specifically as the time interval during which the spatial position is valid. In other words, temporal information is defined by spatial information through a validity period associated with spatial positions. 

Furthermore, we can extract temporal information by recording the order in which spatial positions are observed, rather than recording the detailed validity period of each position. This approach allows us to consider temporal information as a notion of \emph{temporal evolution}, which reflects changes in spatial information or the occurrence of thematic events.

Temporal information encompasses temporal evolution, but the latter efficiently summarizes the concept of time by ordering the observed spatial positions, without necessarily relying on a linear dependency to the real-world time of the phenomenon, which can be enough information in some study cases.

\section{Generic model of an attacking play}
\label{sec:genericmodel}
Taking into account the information described above, we define in this section a new spatio-temporal graph-based model for an \emph{attacking play} in team sports. The concept of attacking play~\citep{ramos2018} refers to the tactical situation in which a team is in possession of the ball and aims to progress toward the opponent's goal. The attacking play, also called \emph{ball possession}, consists of a sequence of actions performed by the players (e.g. passes, runs), with a defined beginning (e.g. a kick-off, intercepting the ball) and end (e.g. scoring, losing the ball). The attacking play represents the fundamental scale for analyzing team sport's phenomena.

In the general framework of ST models, we study entities and the relations that bind them together. In the graph-based model proposed, the entities are of two types: the state of the game (or spatial position), which is the knowledge of relative and absolute spatial information, or the result of the attacking play. The relations are of three types: spatial relations $\mathcal{R}_\text{sp}$ between two states of the game; thematic relations $\mathcal{R}_\text{th}$ between two states of the game too; and result relations $\mathcal{R}_\text{res}$ between a state of the game and a result. 

First, let us define the skeleton graph, in which entities and all possible relations between them are represented. Second, we define how we modelize a given attacking play as a labeled path on the skeleton graph. Note that the skeleton graph is defined \emph{a priori}, i.e. without observing any attacking play.

\subsection{Skeleton graph}
We define the non-directed graph called skeleton graph $\mathcal{K} = (V_\mathcal{K}, E_\mathcal{K})$ as follows, where vertices are entities and edges represent relations. This graph represents all possible relations between the different entities.

\paragraph{Vertices.}
The vertices of $\mathcal{K}$ are of two types: the \emph{spatial vertices} representing the state of the game and the \emph{results vertices} representing how ended the attacking play. We note $V_\text{sp}$, respectively $V_\text{res}$, the sets of theses vertices, thus $V_\mathcal{K} = V_\text{sp} \cup V_\text{res}$. The set of spatial vertices is composed of all the possible couples of relative and absolute information $V_\text{sp} = \mathcal{R}\times\mathcal{A}$. The set of results vertices represents all the possible ways to end a attacking play for the sport at stake, $V_\text{res} = \{\text{res}_1,\ldots,\text{res}_l\}$ for $l\in \mathbb{N}^*$. 

\paragraph{Edges.}
The edges of $\mathcal{K}$ represent all the possible relations between vertices in $V_\mathcal{K}$. The first set of labeled edges $E_\text{sp}$ expresses the spatial relations $\mathcal{R}_\text{sp}$ between two spatial vertices, 
$$E_\text{sp} = \{(\{s_1, s_2\}, \rho_\text{sp}) : s_1, s_2 \in V_\text{sp}, s_1 \neq s_2, \rho_\text{sp} \subseteq \mathcal{R}_\text{sp}\}\,.$$ 
The second set of labeled edges $E_\text{th}$ reflects the thematic relations $\mathcal{R}_\text{th}$ between two spatial vertices too, 
$$E_\text{th} = \{(\{s_1, s_2\}, \rho_\text{th}) : s_1, s_2 \in V_\text{sp}, \rho_\text{th} \subseteq \mathcal{R}_\text{th}\}\,.$$ 
In order to avoid unnecessary multiple graph, we consider the following set of labeled edges 
$$E_\text{sp-th} = \{(\{s_1, s_2\},\rho) : s_1, s_2 \in V_\text{sp}, \rho \subseteq \mathcal{R}_\text{sp} \cup \mathcal{R}_\text{th}\}\,,$$
where the subset $\rho$ is a label indicating the nature of the edge -- which can be spatial and/or thematic relations. Notice that for self-loops $(s,s)$ for $s \in V_\text{st}$, there is no spatial change by definition, thus the label is only related to thematic events.
The third and last set of edges $E_\text{res}$ indicates the possible result relations $\mathcal{R}_\text{res}$ between a spatial vertex and a result vertex, 
$$E_\text{res} = \{\{s, r\} : s \in V_\text{sp}, r \in V_\text{res}\}\,.$$
Eventually, the edges of the skeleton graph are
$$E_\mathcal{K} = E_\text{sp-th} \cup E_\text{res}\,.$$

Notice that, in the generic model, each edge represents a possible relation between two entities, which we assume to be always possible in the general case. For a team sports use-case, some edges could be removed from the skeleton graph if not representing a feasible relation. 

\begin{figure}[ht]
    \centering
\begin{tikzpicture}
    \node (A) at (0, 2) {(0,2),$A_1$};
    \node (B) at (2, 3) {(0,2),$A_2$};
    \node (C) at (4, 2) {(1,1),$A_1$};
    \node (D) at (0, 0) {(1,1),$A_2$};
    \node (E) at (4, 0) {(2,0),$A_1$};
    \node (F) at (2, -1) {(2,0),$A_2$};
    \node (G) at (6,1.6) {Success};
     \node (H) at (-2,1.6) {Failure};
    
    \draw[ultra thick, color=orange] (A) -- (B) ;
    \draw[ultra thick, color=orange] (A) -- (C) ;
    \draw[ultra thick, color=orange] (A) -- (D) ;
    \draw[ultra thick, color=orange] (A) -- (E) ;
    \draw[ultra thick, color=orange] (A) -- (F) ;
    \draw[color=black] (A) -- (G);
    \draw[color=black] (A) -- (H);
   \draw[ultra thick, color=orange] (B) -- (C);
    \draw[ultra thick, color=orange] (B) -- (D);
    \draw[ultra thick, color=orange] (B) -- (E);
    \draw[ultra thick, color=orange] (B) -- (F);
    \draw[color=black] (B) -- (G);
    \draw[color=black] (B) -- (H);
    \draw[ultra thick, color=orange] (C) -- (D);
    \draw[ultra thick, color=orange] (C) -- (E);
    \draw[ultra thick, color=orange] (C) -- (F);
    \draw[color=black] (C) -- (G);
    \draw[color=black] (C) -- (H);
    \draw[ultra thick, color=orange](D) -- (E);
    \draw[ultra thick, color=orange] (D) -- (F);
    \draw[color=black] (D) -- (G);
    \draw[color=black] (D) -- (H);
    \draw[ultra thick, color=orange] (E) -- (F);
    \draw[color=black] (E) -- (G);
    \draw[color=black] (E) -- (H);
    \draw[color=black] (F) -- (G);
    \draw[color=black] (F) -- (H);
    \draw [loop, out=45, in=135, distance=10mm, dashed, ultra thick, color=orange] (A) edge (A) ;
    \draw [loop, out=45, in=135, distance=10mm, dashed, ultra thick,color=orange] (B) edge (B);
    \draw [loop, out=45, in=135, distance=10mm, dashed,ultra thick,color=orange] (C) edge (C);
    \draw [loop, out=45, in=135, distance=-20mm, dashed, ultra thick,color=orange] (D) edge (D);
    \draw [loop, out=45, in=135, distance=-20mm, dashed,ultra thick,color=orange] (E) edge (E);
    \draw [loop, in=45, out =135, distance=-20mm, dashed,ultra thick,color=orange] (F) edge (F);
\end{tikzpicture}
\caption{Example of the skeleton graph for $n=3$, $\mrel = 2$, $\mabs = 2$, and two results vertices. For more readability, we omit the labels. Edges in $E_\text{sp-th}$ for spatial and/or thematic relations are in thick orange, and we specify edges for which there are only thematic labels in dashed thick orange. Edges in $E_\text{res}$ for result relations are in thin black.}
\label{fig:ex_clique}
\end{figure}
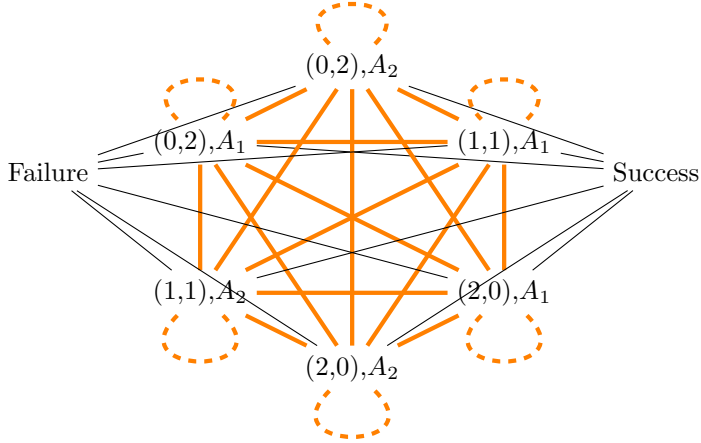

In Figure~\ref{fig:ex_clique}, we provide an example of the skeleton graph for the case of $n=3$ players, $\mrel = 2$ relative zones and $\mabs = 2$ absolute zones. Moreover, we consider the set of result vertices to be $V_\text{res} = \{\textup{Success}, \textup{Failure}\}$. In this case, the result of an attacking play is binary, solely indicating if the attacking play succeeded or failed.

\subsection{Labeled path}

We represent an attacking play as a path in the skeleton graph, augmented with labels on vertices and arcs. Precisely, each spatial vertex, i.e. state of the game, is associated to a given time interval representing temporal information. This time interval $[t_\textup{start}, t_\textup{end}[$ represents the period during which the spatial position associated to the vertex is valid. More over, each arc is labeled representing the nature of the relation, spatial and/or thematic. 

The labeled path is constructed as follows. The starting vertex is the one corresponding to the starting spatial position of the team on the field. As soon as a spatial change and/or a thematic event is detected\footnote{Regarding the implementation, the detection is not continuous, but we can assume that its discretization does not change the construction of the path. Indeed, in practice, the discrete time step for which we obtain the state of the game can always be taken smaller than any thematic change considered.}, we add the vertex corresponding to the new\footnote{The only case where the vertex is identical is when the change is thematic and does not result in a spatial change.} spatial position to the path and the arc's label describing the nature of this change. The time interval on each vertex is labeled in consequence. Notice that during the thematic event, the possible changes of the state of the game are not taken into account, only the difference with the resulting state of the game is expressed. The path ends when the attacking play ends, with an arc from the vertex representing the last spatial position to the corresponding result vertex.

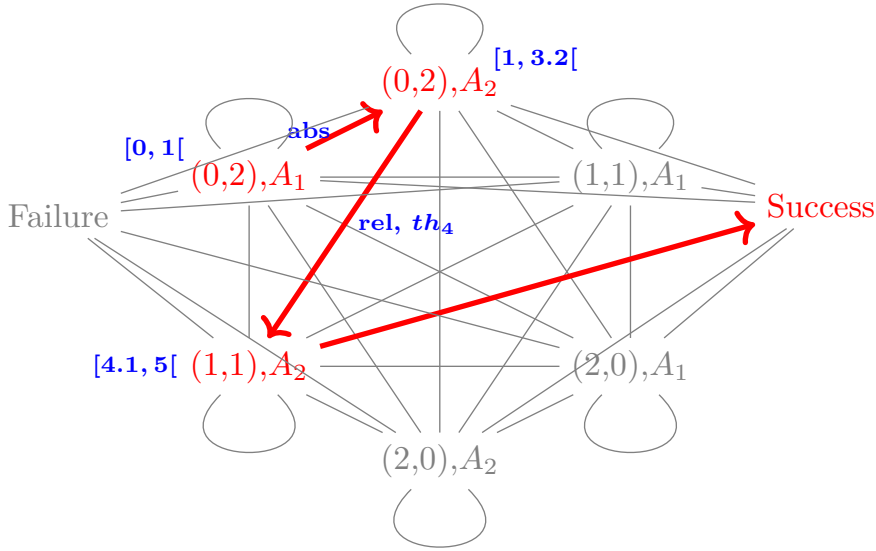
\begin{figure}[ht]
    \centering
    \resizebox{0.8\textwidth}{!}{
\begin{tikzpicture}
    \node[color=red] (A) at (0, 2) {(0,2),$A_1$} ;
    \node[color=red] (B) at (2, 3) {(0,2),$A_2$};
    \node[color=gray] (C) at (4, 2) {(1,1),$A_1$};
    \node[color=red] (D) at (0, 0) {(1,1),$A_2$};
    \node[color=gray] (E) at (4, 0) {(2,0),$A_1$};
    \node[color=gray] (F) at (2, -1) {(2,0),$A_2$};
    \node[color=red]  (G) at (6,1.7) {Success};
     \node[color=gray] (H) at (-2,1.6) {Failure};
   
\node[color=blue] at (-1, 2.3) {\scriptsize \textbf{\mathversion{bold}$[0, 1[$}}; 
\node[color=blue] at (3, 3.25) {\scriptsize \textbf{\mathversion{bold}$[1, 3.2[$}};
\node[color=blue] at (-1.2, 0) {\scriptsize \textbf{\mathversion{bold}$[4.1, 5[$}};
    
    \draw[ultra thick, color=red, ->] (A) -- (B) node[midway, left,color=blue] {\scriptsize \textbf{abs}};
    \draw[color=gray]  (A) -- (C);
    \draw[color=gray]  (A) -- (D);
    \draw[color=gray]  (A) -- (E);
    \draw[color=gray]  (A) -- (F);
     \draw[color=gray]  (A) -- (G);
     \draw[color=gray]  (A) -- (H);
   \draw[color=gray]  (B) -- (C);
    \draw[ultra thick, color=red, ->] (B) -- (D) node[midway, right,color=blue] {\scriptsize \textbf{rel, \mathversion{bold}$th_4$}};
    \draw[color=gray]  (B) -- (E);
    \draw [color=gray] (B) -- (F);
     \draw[color=gray]  (B) -- (G);
     \draw[color=gray]  (B) -- (H);
    \draw [color=gray] (C) -- (D);
    \draw [color=gray] (C) -- (E);
    \draw [color=gray] (C) -- (F);
     \draw[color=gray]  (C) -- (G);
      \draw[color=gray]  (C) -- (H);
    \draw[color=gray]  (D) -- (E);
    \draw[color=gray]  (D) -- (F);
     \draw[ultra thick, color=red, ->]  (D) -- (G) ;
      \draw[color=gray]  (D) -- (H);
    \draw[color=gray]  (E) -- (F);
     \draw[color=gray]  (E) -- (G);
      \draw[color=gray]  (E) -- (H);
      \draw[color=gray]  (F) -- (G);
       \draw[color=gray]  (F) -- (H);
    \draw [loop, out=45, in=135, distance=10mm, color=gray] (A) edge (A);
    \draw [loop, out=45, in=135, distance=10mm, color=gray] (B) edge (B);
    \draw [loop, out=45, in=135, distance=10mm, color=gray] (C) edge (C);
    \draw [loop, out=45, in=135, distance=-20mm, color=gray] (D) edge (D);
    \draw [loop, out=45, in=135, distance=-20mm, color=gray] (E) edge (E);
    \draw [loop, in=45, out =135, distance=-20mm, color=gray] (F) edge (F);
\end{tikzpicture}
    }
\caption{Example of a path on a skeleton graph.}
\label{fig:ex_clique_with_path}
\end{figure}

In Figure~\ref{fig:ex_clique_with_path}, we depict an example of a path on the skeleton graph of the previous example Figure~\ref{fig:ex_clique}. The initial vertex is $((0,2),A_1)$. The attacking play of this example is successful and lasts 5 seconds. The corresponding path is composed of 4 vertices and 3 arcs. The vertices and arcs of the paths are in red, and their respective labels in blue. The first arc represents a change in absolute spatial position, whereas the second arc represents a change in relative spatial position along with a thematic change. Notice that, unlike the first arc, the second arc characterizes a non-instantaneous change due to the thematic event of type $th_4$, which specifically lasts for 0.9 seconds. The third and final arc provides the result of the attacking play.

\subsection{Measures on the model}
\label{subsec:measures}
The model proposed above aims at being a tool to analyze behaviors in team sports. Specifically, given an observed match (i.e. a succession of attacking plays for both teams), we modelize it as a set of labeled paths as presented previously. The goal of this paragraph is to discuss the different possible ways of \emph{analyzing} such outputs of the model, namely to provide some measures on them. 

The analysis of the output model (i.e. a set of attacking plays) is composed of two dimensional aspects. 
The first dimension is the \emph{scale} of the measure: we can consider \emph{local} measures, at the level of a labeled path, or \emph{global} measures, at the level of a graph. The latter graph is the weighted union of several paths selected from the set of attacking plays, thus being a subgraph of the skeleton graph. 
The second dimension is the \emph{nature} of the measure, which can be a \emph{similarity} or a \emph{feature}. On the one hand, similarity relates to a comparison between two mathematical objects (pair of paths or pair of subgraphs in our case), such as a distance, meaning that the comparison is pairwise. On the other hand, features are absolute indicators of a given mathematical object, absolute in the sense that their values does not depend on other objects. 

The challenge of such analysis is to find a way to extract relevant absolute indicators (for features) and quantify pertinently the difference between two outputs (for similarities). For instance, at the local scale, defining a metric comparing two paths is complex because these two paths have not necessarily the same length, and they carry high-level information (absolute and relative position on the vertices, type of events on the arcs etc.). This work requires both a rigorous definition of the metric (function that satisfies non-negativity, equality to zero for the same object, symmetry, triangle inequality) and quantitative decisions based on the particularity of the team sport in question. Notice that in Section~\ref{sec:experiments}, we use generic features in graph theory (e.g. path length, density) and specific features defined according to the hypothesis at stake (see Subsections~\ref{subsec:defense_constraints} and~\ref{subsec:pedagogy_constraints}), both at local and global scales, to analyze the results of the experiments on rugby. The use of new measures, more adapted to the specificity of the model, is discussed in Section~\ref{sec:discussion} and is ongoing work.

\section{Application to rugby}
\label{sec:appli_rugby}

In this section, we illustrate the model on the game of rugby. Applying this model to a team sports amounts to choose the parameters of the model, which are the partitions in $\mabs$ absolute zones and $\mrel$ relative zones, the thematic events considered and the result vertices. These choices are led by the specificity of the sport. The aim of rugby is to make the ball progress through the field toward the try line, which makes us naturally consider absolute zones parallel to the try line. Moreover, any player ahead of the ball carrier is offside and cannot receive the ball, thus we do not specifically express the number of players ahead in the relative zones because it is a temporary situation.

In the present modelization -- motivated by experimental protocols presented later -- we consider a rugby small-sided game, namely a rugby game with a reduced number of players in a smaller playing area. Specifically, the field is a rectangle (dimensions are detailed below), where one of the smallest edge is the try line. The aim of the attacking team, composed of $n=6$ players, is to bring the ball to the try line. The results considered in the application are whether the attacking play succeeds (``Try'' vertex) or not (``Failure'' vertex).

\subsection{Parameters of the model}

Let us first define the absolute and relative spatial information we consider for rugby small-sided game. 

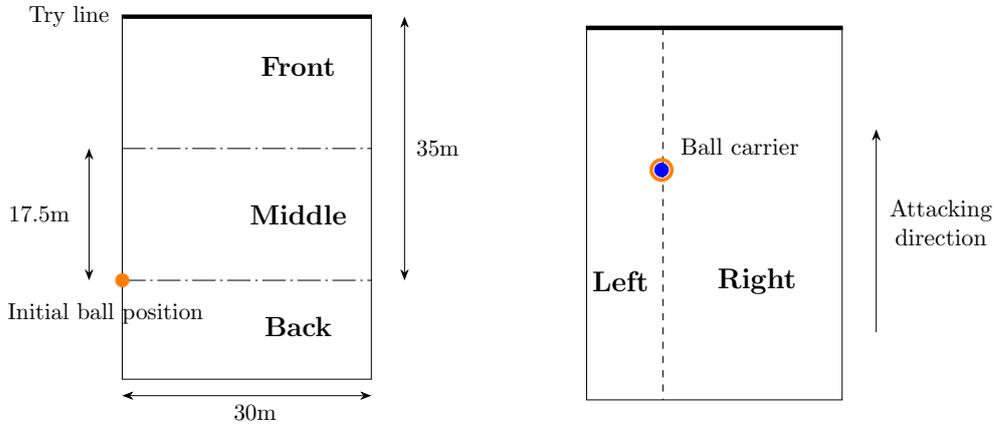
\begin{figure}[h!]
    \centering
    \begin{minipage}{0.49\textwidth}
        \centering
        \resizebox{0.85\textwidth}{!}{%
        \begin{circuitikz}
        \tikzstyle{every node}=[font=\LARGE]
        \draw (5,5) -- (5,10.5) -- (8.75,10.5) -- (8.75,5)  -- cycle ; 
        \draw [dash pattern=on 10pt off 2pt on 1pt off 2pt] (5,6.5) -- (8.75,6.5);
        \draw [dash pattern=on 10pt off 2pt on 1pt off 2pt] (5,8.5) -- (8.75,8.5);
        \draw [<->, >=Stealth] (9.25,6.5) -- (9.25,10.5);
        \draw [<->, >=Stealth] (4.5,6.5) -- (4.5,8.5);
        \draw [<->, >=Stealth] (5,4.75) -- (8.75,4.75);
        \node [font=\normalsize] at (9.75,8.5) {35m};
        \node [font=\normalsize] at (3.75,7.5) {17.5m};
        \node [font=\normalsize] at (7,4.5) {30m};
        
        \draw [color=black, line width=1.9pt] (4.99,10.5) -- (8.76,10.5);
        \node [font=\normalsize] at (4.2,10.5) {Try line};
        \node [font=\large] at (7.65,9.75) {\textbf{Front}};
        \node [font=\large] at (7.65,7.5) {\textbf{Middle}};
        \node [font=\large] at (7.65,5.8) {\textbf{Back}};
       
        \draw [color=orange, fill=orange, line width=1.5pt] (5,6.5) circle (0.08cm);
        \node [font=\normalsize] at (4.75,6) {Initial ball position};
        
        \end{circuitikz}
        }%
    \end{minipage}%
    \begin{minipage}{0.49\textwidth}
        \centering
        \resizebox{0.76\textwidth}{!}{%
        \begin{circuitikz}
        \tikzstyle{every node}=[font=\LARGE]
        \draw (5,5) -- (5,10.5) -- (8.75,10.5) -- (8.75,5)  -- cycle ;
        \draw [color=black, line width=1.9pt] (4.99,10.5) -- (8.76,10.5);
        \node [font=\large] at (5.5,6.75) {\textbf{Left}};
        \node [font=\large] at (7.5,6.75) {\textbf{Right}};
        \draw [dashed] (6.12,10.5) -- (6.12,5);
        
        \draw [color=blue, fill=blue, line width=0.2pt] (6.1,8.4) circle (0.1cm); 
        \draw [color=orange, fill=none, line width=1.5pt] (6.1,8.4) circle (0.15cm);
        \node [font=\normalsize] at (7.25,8.75) {Ball carrier};
        
        \draw [->, >=Stealth] (9.25,6) -- (9.25,9);
        \node [font=\normalsize] at (10.2,7.8) {Attacking};
        \node [font=\normalsize] at (10.2,7.4) {direction};
    
        \end{circuitikz}
        }%
    \end{minipage}
    \caption{Absolute zones (on the left) and relative zones (on the right) for rugby.}
    \label{fig:abs_and_rel_zones_rugby} 
\end{figure}

\paragraph{Absolute zones.} We decompose the field into 3 absolute zones: ``Back'', ``Middle'' and ``Front'', i.e. 
$$ \mathcal{A} = \{\textup{Back},\textup{Middle},\textup{Front}\}\,.$$
The absolute zones are depicted in the left side of Figure~\ref{fig:abs_and_rel_zones_rugby}.

\paragraph{Relative zones.} For a given position $pos(t)$, we divide the field into 2 relative zones ``Left'' and ``Right'', that are the zones to the left and to the right of the line passing through $pos(t)$ and perpendicular to the try line. Thus, we consider 
$$\mathcal{R} = \{(l,r) : l+r=5,\, l, r\in\mathbb{N}\}\,.$$
The absolute zones are depicted in the right side of Figure~\ref{fig:abs_and_rel_zones_rugby}.

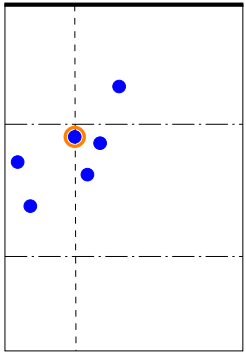
\begin{figure}[h!]
    \centering
        \resizebox{0.22\textwidth}{!}{%
        \begin{circuitikz}
        \tikzstyle{every node}=[font=\LARGE]
        \draw (5,5) -- (5,10.5) -- (8.75,10.5) -- (8.75,5)  -- cycle ; 
        \draw[dash pattern=on 10pt off 2pt on 1pt off 2pt] (5,6.5) -- (8.75,6.5);
        \draw[dash pattern=on 10pt off 2pt on 1pt off 2pt] (5,8.6) -- (8.75,8.6);
        
        \draw [color=black, line width=1.9pt] (4.99,10.5) -- (8.76,10.5);
        
        \draw [dashed] (6.1,10.5) -- (6.12,5);
        \draw [color=blue, fill=blue, line width=0.2pt] (6.1,8.4) circle (0.1cm); 
        \draw [color=orange, fill=none, line width=1.5pt] (6.1,8.4) circle (0.15cm);

        \draw [color=blue, fill=blue, line width=0.2pt] (5.2,8) circle (0.1cm);
        \draw [color=blue, fill=blue, line width=0.2pt] (5.4,7.3) circle (0.1cm);
        \draw [color=blue, fill=blue, line width=0.2pt] (6.5,8.3) circle (0.1cm);
        \draw [color=blue, fill=blue, line width=0.2pt] (6.8,9.2) circle (0.1cm);
        \draw [color=blue, fill=blue, line width=0.2pt] (6.3,7.8) circle (0.1cm);
        \end{circuitikz}
        }%
    \caption{Example of a spatial position at a given instant time, with the 6 players (blue dots) and among them the ball carrier (circled in orange). The spatial information related to this position is ($(2,3)$, Middle).}
    \label{fig:example_spatial_position_rugby} 
\end{figure}

In Figure~\ref{fig:example_spatial_position_rugby}, we provide an example of the spatial information encoded by our model with the spatial partitioning presented above. 

\begin{figure}[h!]
\centering
\begin{tikzpicture}
    \node[circle, draw] at (2,9) (C) {};
    \node at (2,9) {$r$};
    \node[circle, draw] at (0,8) (D) {};
    \node[circle, draw] at (4,8) (E) {};
    \node[circle, draw] at (-1,7) (F) {};
    \node[circle, draw] at (1,7) (G) {};
    \node[circle, draw, fill=blue] at (3,7) (H) {};
    \node[circle, draw] at (5,7) (I) {};

    \draw[thick, color=black] (C) -- (D) node[midway, left] {\small hand};
    \draw[thick, color=black] (C) -- (E) node[midway, right] {\small kick};
    \draw[thick, color=black] (D) -- (F) node[midway, left] {\small no contact};
    \draw[thick, color=black] (D) -- (G) node[midway, right] {\small contact};
    \draw[thick, color=black] (E) -- (H) node[midway, left] {\small diagonal};
    \draw[thick, color=black] (E) -- (I) node[midway, right] {\small straight};
\end{tikzpicture}
\caption{Tree of thematic labels in rugby.}
\label{fig:rugby_thematic_labels}
\end{figure}
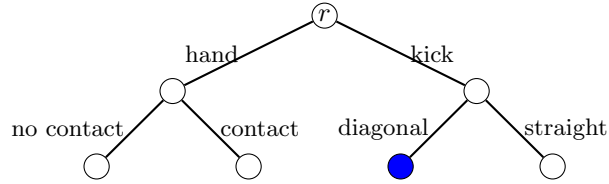

Let us next present the thematic events considered for this application of rugby. 
\paragraph{Thematic events.} We consider thematic events as passes of different nature. A pass can be done either with the hand or with the foot. Regarding a hand pass, it could be performed with or without a \emph{contact}. Indeed, tackling is prohibited in the rugby small-sided game considered: instead, when the ball carrier is touched (with two hands) by an opponent, he has to either place the ball on the ground or make a hand pass (which characterizes a pass with a contact). For a kick pass, it can be either straight toward the try line or in diagonal.
Specifically, we define the set of thematic labels $\mathcal{T}$ as the set of leaves of the tree Figure~\ref{fig:rugby_thematic_labels}. The label corresponding to a leaf is the semantic information present on the path from the root $r$ to itself. 

For instance, the blue leaf represents the thematic label ``kick diagonal''. 

\subsection{An example of attacking play} 

\begin{figure}[ht]
    \centering
    \resizebox{\textwidth}{!}{
    \begin{minipage}{0.26\textwidth}
        \centering
        \begin{circuitikz}
        \tikzstyle{every node}=[font=\LARGE]
        \draw (5,5) -- (5,10.5) -- (8.75,10.5) -- (8.75,5)  -- cycle ; 
        \draw[dash pattern=on 10pt off 2pt on 1pt off 2pt](5,6.5) -- (8.75,6.5);
        \draw[dash pattern=on 10pt off 2pt on 1pt off 2pt] (5,8.6) -- (8.75,8.6);
        
        \draw [color=black, line width=1.9pt] (4.99,10.5) -- (8.76,10.5);
        
        \draw [dashed] (6.1,10.5) -- (6.1,5);
        \draw [color=blue, fill=blue, line width=0.2pt] (6.1,8.4) circle (0.1cm); 
        \draw [color=orange, fill=none, line width=1.5pt] (6.1,8.4) circle (0.15cm);
        \draw [color=blue, fill=blue, line width=0.2pt] (5.2,8) circle (0.1cm);
        \draw [color=blue, fill=blue, line width=0.2pt] (5.9,7.5) circle (0.1cm);
        \draw [color=blue, fill=blue, line width=0.2pt] (6.5,8.3) circle (0.1cm);
        \draw [color=blue, fill=blue, line width=0.2pt] (6.8,9.2) circle (0.1cm);
        \draw [color=blue, fill=blue, line width=0.2pt] (6.3,7.8) circle (0.1cm);
        
        \node [font=\large] at (6.9, 4.7) {$t = t_0$};
        \end{circuitikz}
    \end{minipage}
    \begin{minipage}{0.26\textwidth}
        \centering
        \begin{circuitikz}
        \tikzstyle{every node}=[font=\LARGE]
        \draw (5,5) -- (5,10.5) -- (8.75,10.5) -- (8.75,5)  -- cycle ; 
        \draw[dash pattern=on 10pt off 2pt on 1pt off 2pt] (5,6.5) -- (8.75,6.5);
        \draw[dash pattern=on 10pt off 2pt on 1pt off 2pt](5,8.6) -- (8.75,8.6);
        
        \draw [color=black, line width=1.9pt] (4.99,10.5) -- (8.76,10.5);
        
        \draw [dashed] (6.5,10.5) -- (6.5,5);
        \draw [color=blue, fill=blue, line width=0.2pt] (6.5,8.9) circle (0.1cm); 
        \draw [color=orange, fill=none, line width=1.5pt] (6.5,8.9) circle (0.15cm);
        \draw [color=blue, fill=blue, line width=0.2pt] (5.3,8.1) circle (0.1cm);
        \draw [color=blue, fill=blue, line width=0.2pt] (6.7,7.9) circle (0.1cm);
        \draw [color=blue, fill=blue, line width=0.2pt] (6.9,7.9) circle (0.1cm);
        \draw [color=blue, fill=blue, line width=0.2pt] (6.8,8.3) circle (0.1cm);
        \draw [color=blue, fill=blue, line width=0.2pt] (7,9) circle (0.1cm);
        
        \node [font=\large] at (6.9, 4.7) {$t = t_0 + \delta$};
        \end{circuitikz}
    \end{minipage}
    \begin{minipage}{0.26\textwidth}
        \centering
        \begin{circuitikz}
        \tikzstyle{every node}=[font=\LARGE]
        \draw (5,5) -- (5,10.5) -- (8.75,10.5) -- (8.75,5)  -- cycle ; 
        \draw[dash pattern=on 10pt off 2pt on 1pt off 2pt](5,6.5) -- (8.75,6.5);
        \draw[dash pattern=on 10pt off 2pt on 1pt off 2pt] (5,8.6) -- (8.75,8.6);
        
        \draw [color=black, line width=1.9pt] (4.99,10.5) -- (8.76,10.5);
        
        \draw [dashed] (6.7,10.5) -- (6.7,5);
        \draw [color=blue, fill=blue, line width=0.2pt] (6.7,9.3) circle (0.1cm); 
        \draw [color=orange, fill=none, line width=1.5pt] (6.7,9.3) circle (0.15cm);
        \draw [color=blue, fill=blue, line width=0.2pt] (5.7,8.2) circle (0.1cm);
        \draw [color=blue, fill=blue, line width=0.2pt] (6.85,8.2) circle (0.1cm);
        \draw [color=blue, fill=blue, line width=0.2pt] (7.3,7.8) circle (0.1cm);
        \draw [color=blue, fill=blue, line width=0.2pt] (7.2,8.3) circle (0.1cm);
        \draw [color=blue, fill=blue, line width=0.2pt] (7.6,8.9) circle (0.1cm);
        
        \node [font=\large] at (6.9, 4.7) {$t = t_0 + 2\delta$};
        \end{circuitikz}
    \end{minipage}
    
    \begin{minipage}{0.26\textwidth}
        \centering
        \begin{circuitikz}
        \tikzstyle{every node}=[font=\LARGE]
        \draw (5,5) -- (5,10.5) -- (8.75,10.5) -- (8.75,5)  -- cycle ; 
        
        \draw [color=black, line width=1.9pt] (4.99,10.5) -- (8.76,10.5);
        
        \draw[->] (6.9,9.6) to (7.8,9.325); 
        \draw [color=blue, fill=blue, line width=0.2pt] (6.9,9.6) circle (0.1cm); 
        \draw [color=orange, fill=orange, line width=1.5pt] (7.3,9.49) circle (0.05cm);
        
        \draw [color=blue, fill=blue, line width=0.2pt] (6.5,8.4) circle (0.1cm);
        \draw [color=blue, fill=blue, line width=0.2pt] (7,8.6) circle (0.1cm);
        \draw [color=blue, fill=blue, line width=0.2pt] (7.6,8.3) circle (0.1cm);
        \draw [color=blue, fill=blue, line width=0.2pt] (7.2,8.7) circle (0.1cm);
        \draw [color=blue, fill=blue, line width=0.2pt] (7.9,9.3) circle (0.1cm);

        \node [font=\large] at (6.9, 4.7) {$t = t_0 + 3\delta$};
        \end{circuitikz}
    \end{minipage}
    \begin{minipage}{0.26\textwidth}
        \centering
        
        \begin{circuitikz}
        \tikzstyle{every node}=[font=\LARGE]
        \draw (5,5) -- (5,10.5) -- (8.75,10.5) -- (8.75,5)  -- cycle ; 
        \draw[dash pattern=on 10pt off 2pt on 1pt off 2pt] (5,6.5) -- (8.75,6.5);
        \draw[dash pattern=on 10pt off 2pt on 1pt off 2pt](5,8.6) -- (8.75,8.6);
        
        \draw [color=black, line width=1.9pt] (4.99,10.5) -- (8.76,10.5);
        
        \draw [dashed] (7.9,10.5) -- (7.9,5);
        \draw [color=blue, fill=blue, line width=0.2pt] (7.9,9.7) circle (0.1cm); 
        \draw [color=orange, fill=none, line width=1.5pt] (7.9,9.7) circle (0.15cm);
        
        \draw [color=blue, fill=blue, line width=0.2pt] (6.5,9.1) circle (0.1cm);
        \draw [color=blue, fill=blue, line width=0.2pt] (7.2,8.8) circle (0.1cm);
        \draw [color=blue, fill=blue, line width=0.2pt] (8.1,9) circle (0.1cm);
        \draw [color=blue, fill=blue, line width=0.2pt] (7.5,9) circle (0.1cm);
        \draw [color=blue, fill=blue, line width=0.2pt] (7,9.5) circle (0.1cm);
        
        \node [font=\large] at (6.9, 4.7) {$t = t_0 + 4\delta$};
        \end{circuitikz}
    \end{minipage}
    \begin{minipage}{0.26\textwidth}
        \centering
        \begin{circuitikz}
        \tikzstyle{every node}=[font=\LARGE]
        \draw (5,5) -- (5,10.5) -- (8.75,10.5) -- (8.75,5)  -- cycle ; 
        \draw[dash pattern=on 10pt off 2pt on 1pt off 2pt] (5,6.5) -- (8.75,6.5);
        \draw[dash pattern=on 10pt off 2pt on 1pt off 2pt] (5,8.6) -- (8.75,8.6);
        
        \draw [color=black, line width=1.9pt] (4.99,10.5) -- (8.76,10.5);
        
        \draw [dashed] (8,10.5) -- (8,5);
        \draw [color=blue, fill=blue, line width=0.2pt] (8,10.35) circle (0.1cm); 
        \draw [color=orange, fill=none, line width=1.5pt] (8,10.35) circle (0.15cm);
        
        \draw [color=blue, fill=blue, line width=0.2pt] (6.1,9.4) circle (0.1cm);
        \draw [color=blue, fill=blue, line width=0.2pt] (7.2,9.2) circle (0.1cm);
        \draw [color=blue, fill=blue, line width=0.2pt] (8.4,9.7) circle (0.1cm);
        \draw [color=blue, fill=blue, line width=0.2pt] (7.8,9.4) circle (0.1cm);
        \draw [color=blue, fill=blue, line width=0.2pt] (6.7,9.9) circle (0.1cm); carrier
        
        \node [font=\large] at (6.9, 4.7) {$t = t_0 + 5\delta$};
        \end{circuitikz}
    \end{minipage}
    }
    \caption{Example of an attacking play for rugby small-sided game ($n = 6$, $\mrel = 2$, $\mabs = 3$, zoning of Figure~\ref{fig:abs_and_rel_zones_rugby}). The game is observed every $\delta$ seconds. The players are blue dots, and the ball is symbolized by an orange circle, rounding the player carrying it when there is no pass. A pass is symbolized by an arrow (see frame at time $t_0 + 3\delta)$.}
    \label{fig:example_possession}
\end{figure}
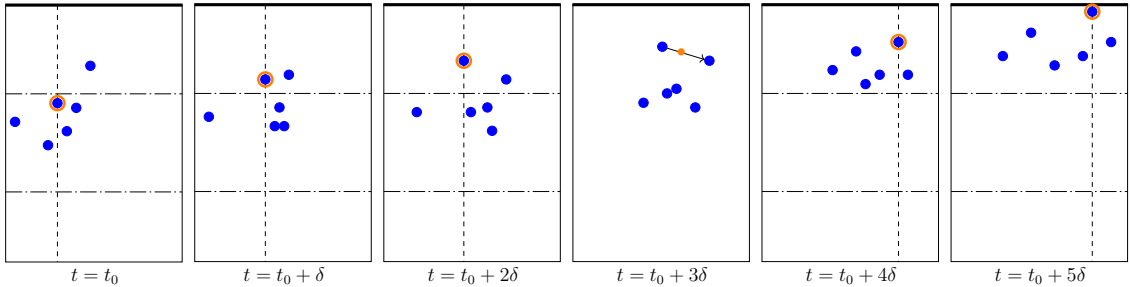

\begin{figure}[!ht]
    \centering
\resizebox{1.\textwidth}{!}{%
\begin{circuitikz}
\tikzstyle{every node}=[font=\LARGE]

\node (tikzmaker) at (0,0) {\includegraphics[width=17cm]{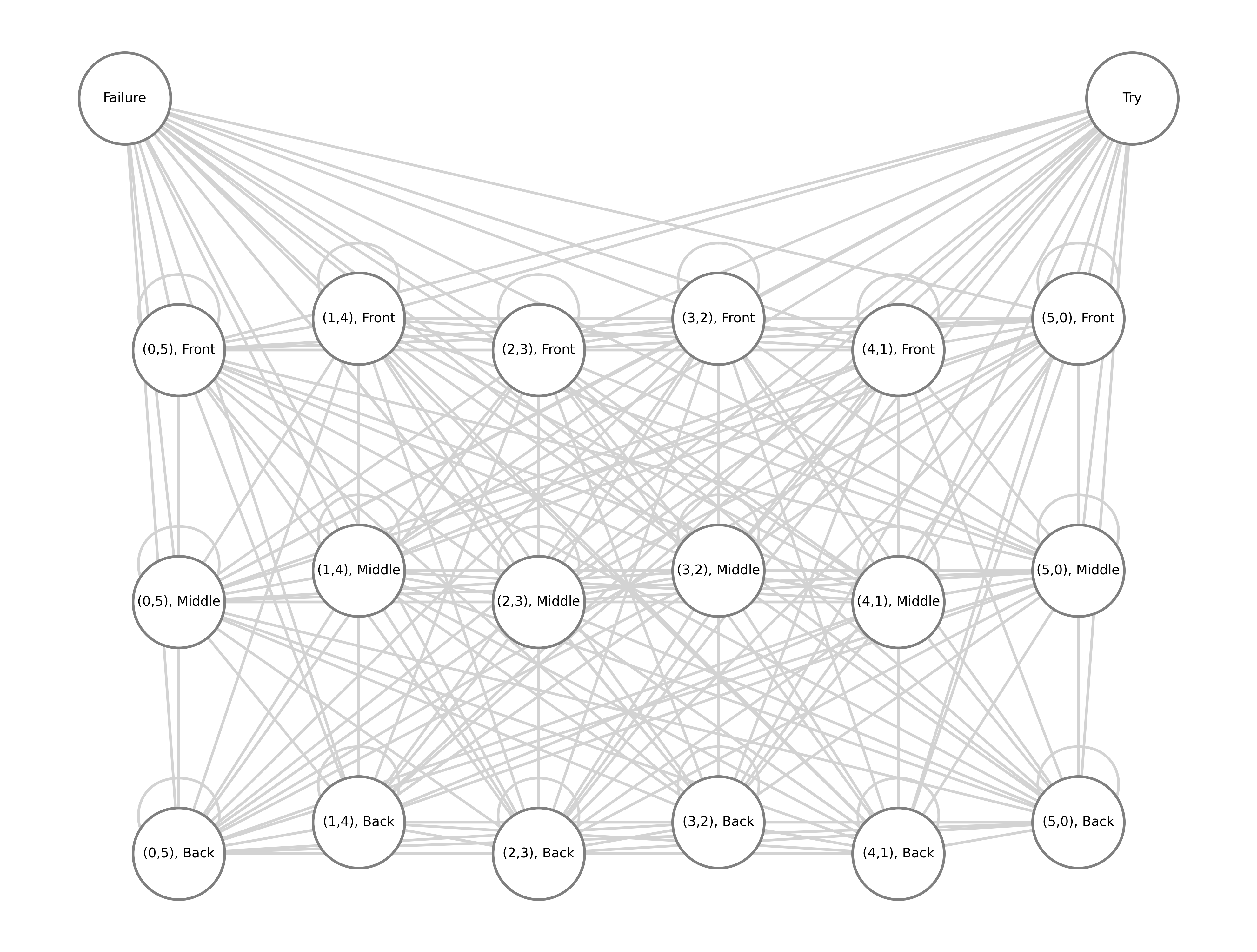}};

\draw [color=red, fill=none, line width=2pt] (-1.22,-1.7) circle (0.62cm); 
\draw [color=red, fill=none, line width=2pt] (-3.65,2.14) circle (0.62cm); 
\draw [color=red, fill=none, line width=2pt] (3.65,1.7) circle (0.62cm); 
\draw [color=red, fill=none, line width=2pt] (6.82,5.12) circle (0.62cm); 

\draw[->, color=red, line width=2pt] (-1.5,-1.18) to (-3.35,1.55);
\draw[->, color=red, line width=2pt] (-3,2.14) to (3,1.7);
\draw[->, color=red, line width=2pt] (3.86,2.3) to (6.3,4.75);

\node[color = blue] at (-1.5,-2.6) {\small \textbf{\mathversion{bold}$[t_0, t_0 + \delta [$}}; 
\node[color = blue] at (-4.2,3)  {\small \textbf{\mathversion{bold}$[t_0 + \delta, t_0 + 3\delta [$}}; 
\node[color = blue] at  (3.8,0.8) {\small \textbf{\mathversion{bold}$[t_0 + 4\delta, t_0 + 5\delta [$}}; 

\node[color = blue] at (-1.4,0) {\small \textbf{abs, rel}}; 
\node[color = blue] at (-0.9,2.5) {\small \textbf{rel, hand no contact}}; 

\end{circuitikz}
}%

\caption{Skeleton graph of rugby small-sided game. In red, we draw the path corresponding to the attacking play of Figure~\ref{fig:example_possession}.}
\label{fig:skeleton_graph_example_possession}
\end{figure}
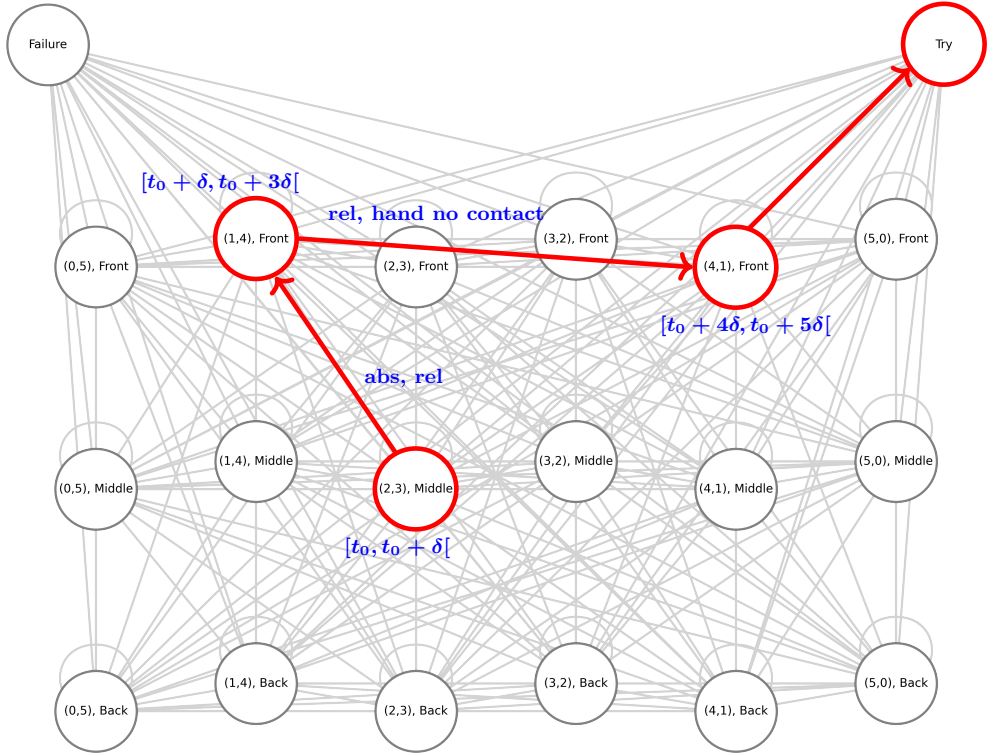

We present in Figure~\ref{fig:example_possession} an example of an attacking play starting at time $t_0$, with a discrete time step detection of $\delta$. In Figure~\ref{fig:skeleton_graph_example_possession}, we depict the skeleton graph for the specific parameters of the model for the rugby application and the path corresponding to this attacking play. 

Note that the vertices of the skeleton graph are arranged in such a way that they reflect the topology of the field. Indeed, all the vertices corresponding to the same absolute zone (``Back'', ``Middle'', and ``Front'') are placed in the same row, which is positioned according to the absolute zone it represents (respectively at the bottom, in the middle, and at the top of the graph). Moreover, the vertices corresponding to the same relative zone (six relative zones in total) are placed in the same column, which is positioned according to the relative zone it represents. For instance, relative zones with a large number of players in the ``Right'' zone are in the left columns, whereas, relative zones with few players in the ``Right'' zone, and thus most of the players in the ``Left'' zone, are in the right columns.

\subsection{Visualization of the model output}
\label{subsec:visualisation}
In what follows, we provide examples of visualization of the model output reflecting different aspects for a given set of attacking plays.  

\begin{figure}[h!]
    \centering
    \subfigure[Set of attacking plays $A$.]{\includegraphics[width=0.49\textwidth]{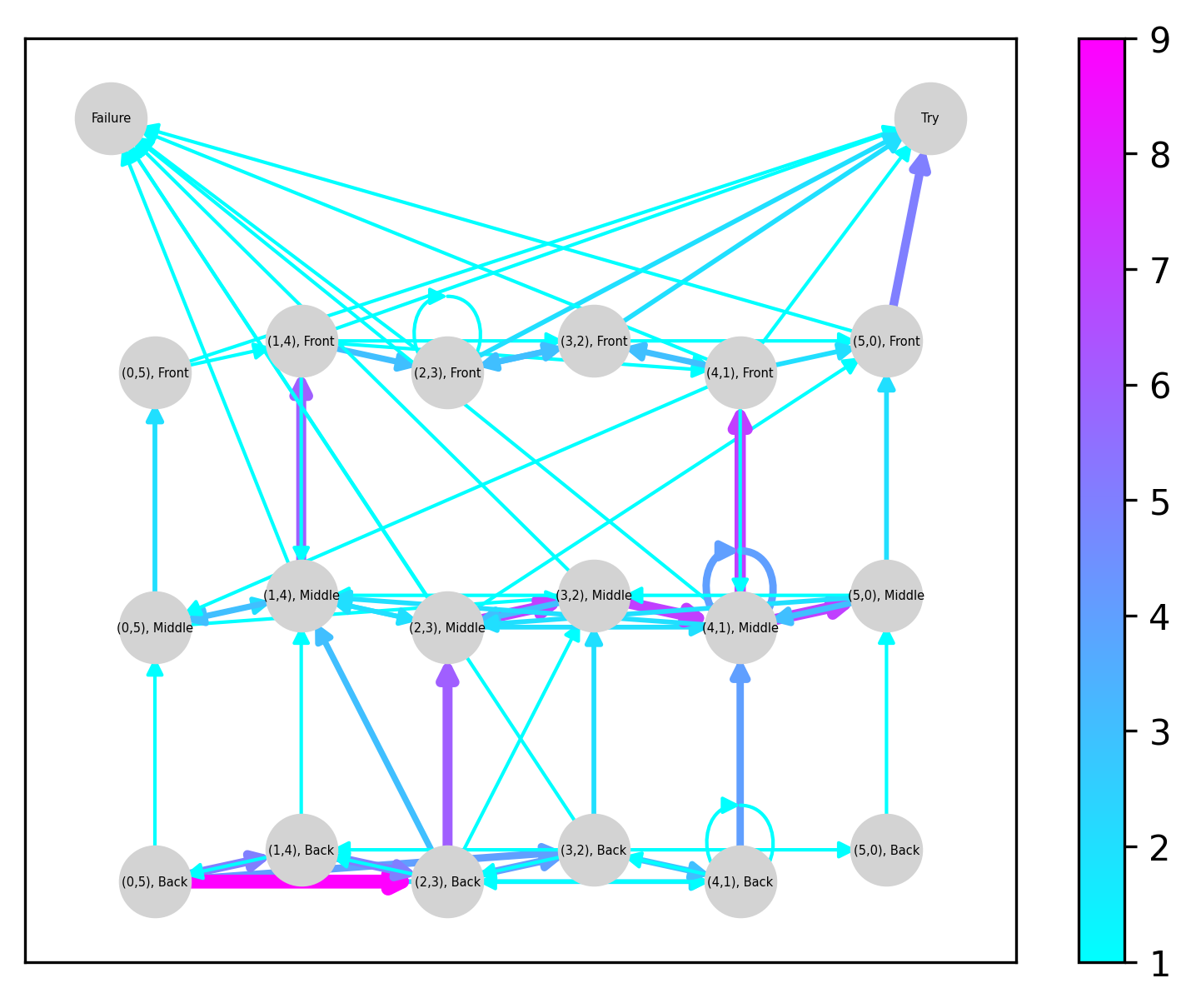}\label{fig:pedagogies_occurence_centrality_edges_L_Kick}}
    \subfigure[Set of attacking plays $B$.]{\includegraphics[width=0.49\textwidth]{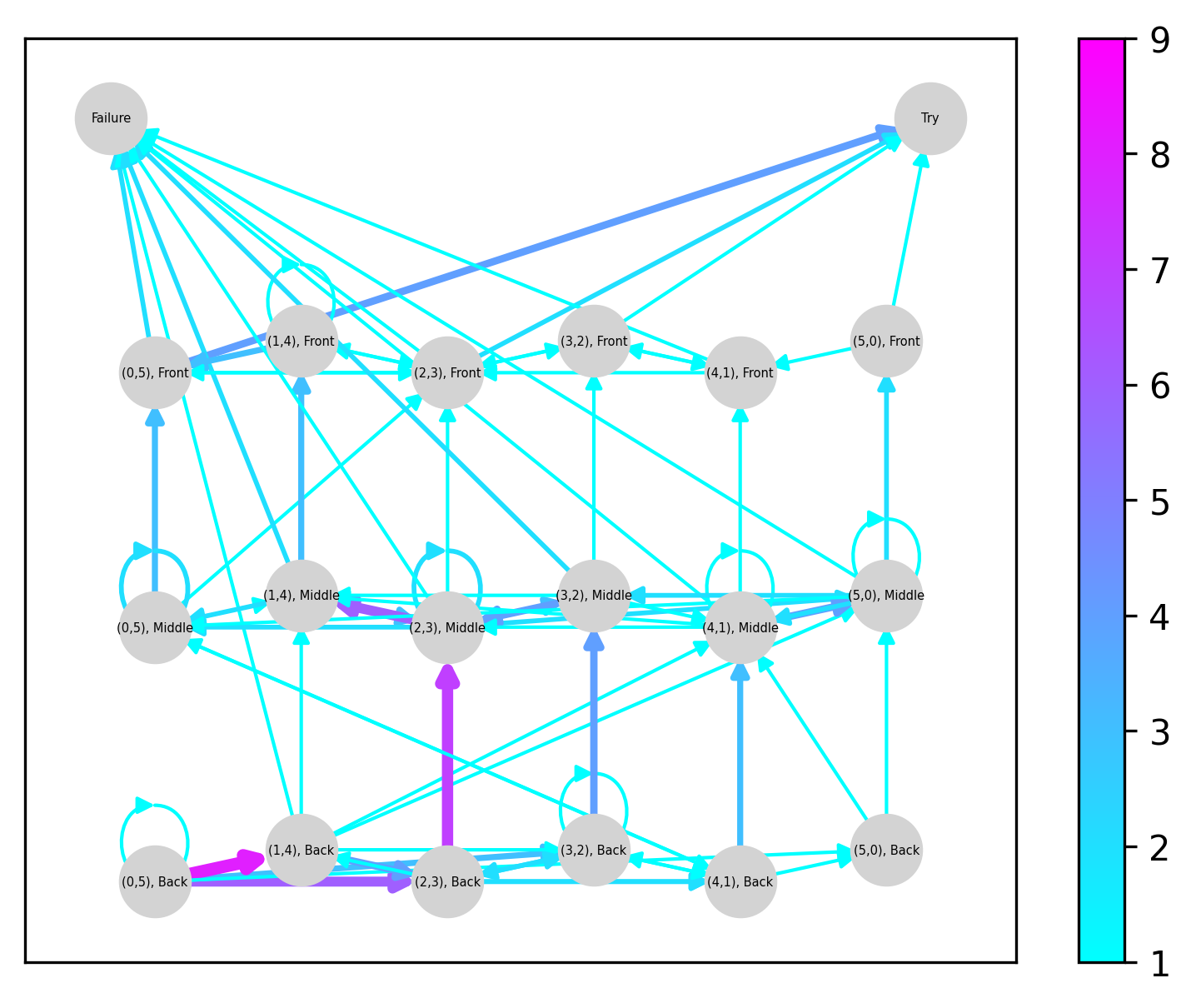}\label{fig:pedagogies_occurence_centrality_edges_NL_Kick}}
    \caption{Occurrence centrality edges for two sets $A$ and $B$ of attacking plays.}
\end{figure}

In Figures~\ref{fig:pedagogies_occurence_centrality_edges_L_Kick} and~\ref{fig:pedagogies_occurence_centrality_edges_NL_Kick}, we display the occurrence centrality edges of two different sets of attacking plays\footnote{These two sets come from the experimental protocol defined in Section~\ref{sec:experiments}. However, at this time of the paper, there is no need to provide the details of what represents each set.}, set $A$ et set $B$. 
The vertices are those of the skeleton graph (see Figure~\ref{fig:skeleton_graph_example_possession}) and the arcs are weighted (notified by color and width) by the number of time the paths representing the attacking plays considered contain them. Notice that, in this example, each attacking play starts from the same initial spatial position on the field corresponding to the vertex ``(0,5), Back'' at the bottom left.

\begin{figure}[h!]
    \centering
    \subfigure[Set of attacking plays $A$.]{\includegraphics[width=0.49\textwidth]{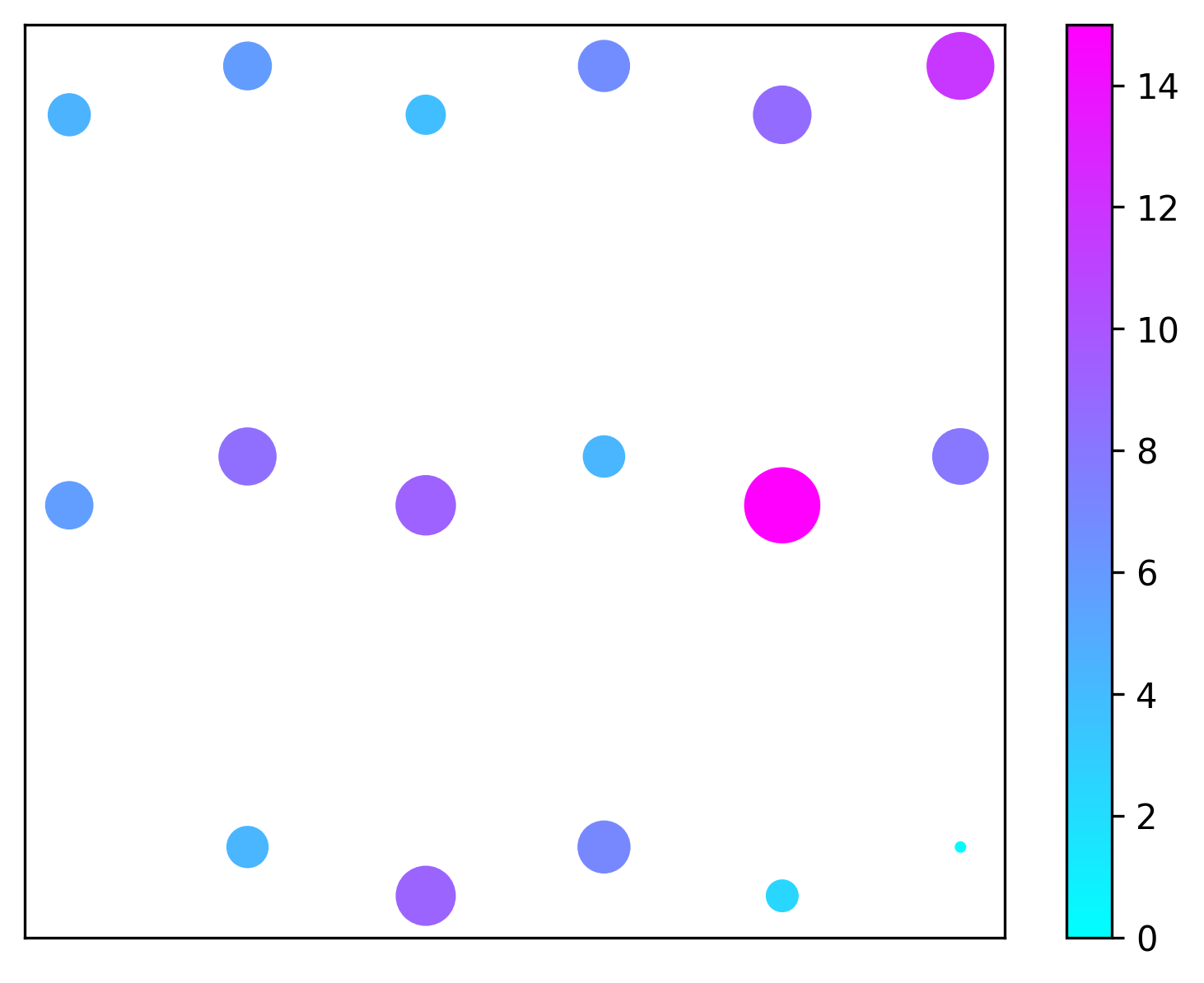}\label{fig:pedagogies_timespent_vertices_L_Kick}}
    \subfigure[Set of attacking plays $B$.]{\includegraphics[width=0.49\textwidth]{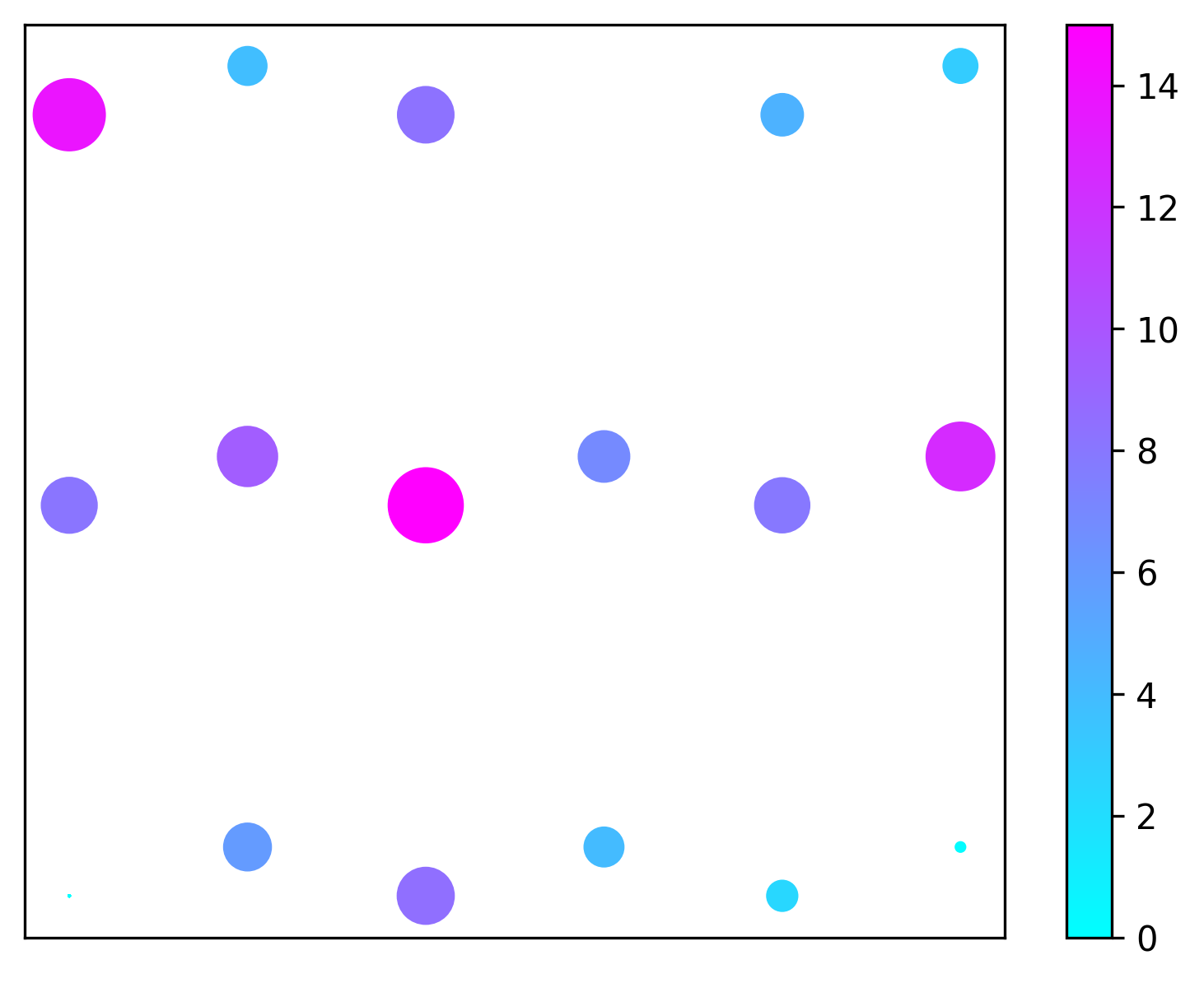}\label{fig:pedagogies_timespent_vertices_NL_Kick}}
    \caption{Time spent (in seconds) on vertices for two sets $A$ and $B$ of attacking plays. The size of a vertex is the associated time spent multiplied by 30 for more readability.}
\end{figure}

In previous visualization, we focus on the changes of spatial states. In Figures~\ref{fig:pedagogies_timespent_vertices_L_Kick} and~\ref{fig:pedagogies_timespent_vertices_NL_Kick}, we represent -- for the same two set of attacking plays -- another information provided by the model: the time spent on vertices. Each vertex of the skeleton graph is weighted (notified by color and size) by the sum of time spent on it for each path of the set of attacking plays considered. Note that we display the vertices of the skeleton graph without the result vertices because no time period is associated with them (neither with the initial vertex on bottom left, which represents the initial position on the field and from which the attacking play starts when this position is left).

\section{Experiments}
\label{sec:experiments}

\subsection{Constraints and hypotheses}

An experimental protocol was designed to evaluate the effects of two constraints on team coordination patterns during a small-sided game representative of a typical rugby situation (i.e. a post-ruck scenario). The game involved 6 attackers versus 6 defenders on a field measuring 30 meters in width and 35 meters in length, starting from a static position near the sideline (see Figure~\ref{fig:abs_and_rel_zones_rugby}). Attackers executed repeated attacking plays, each initiated by a first pass and ending either with a try or a loss of ball possession. The two constraints studied, defensive scenario and pedagogy, are detailed below together with the related hypotheses.

On the one hand, attackers were confronted with different defensive scenarios, designed by modifying the initial positions of the defenders, to invite distinct tactical solutions for attackers:
\begin{itemize}
    \item Tight scenario: the first line of defenders was compactly grouped near the ruck, conceding wide spaces on the wing. 
    \begin{hypothesis_box}{Hypothesis H1.A}
    The tight scenario invites the attackers to pass the ball laterally to bypass the defenders and reach the opposite wing.
    \end{hypothesis_box}
    \item Open scenario: the first line of defenders was spread across the width of the field, creating large gaps between them.
    \begin{hypothesis_box}{Hypothesis H1.B} 
    The open scenario invites the attackers to carry the ball to cross the gaps and break through the defense.
    \end{hypothesis_box}
    \item Kick scenario: the first line of defenders was more effectively covering the front line (i.e. no wide spaces or large gaps), but leaving space behind them.\footnote{For the kick scenario, the front line of defenders was composed of 5 defenders and only 1 defender was in the second line, leaving space behind the defense. Conversely, in the tight scenario, the front line was composed of 4 defenders and the second line was composed of 2 defenders, leaving space in the wing of the defense.}
    \begin{hypothesis_box}{Hypothesis H1.C} The kick scenario invites attackers to use kick passes to pass over the defense and exploit the space behind them.
    \end{hypothesis_box}
\end{itemize}

On the other hand, two distinct pedagogical approaches were implemented: a \emph{linear pedagogy}, and a \emph{non-linear pedagogy}. The linear pedagogy consisted in a prescriptive approach of learning, in which the coach explicitly instructed players about the optimal tactical solution to apply in each defensive scenario, aiming to reproduce predefined patterns of play. By contrast, in the non-linear pedagogy condition, the coach promoted self-organization by providing only a few collective organization principles that remained the same regardless of the defensive scenario. In other words, in the linear pedagogy, specific individual instructions were provided to each player, while in the non-linear pedagogy, the instructions focused on collective organization principles given to all players. Specifically, these principles were: 1) providing three supports for the ball carrier, 2) playing in the gaps, and 3) utilizing open space. While linear pedagogy relies on the assumption that players need to have the same plan in mind (i.e. share knowledge and mental representations about the team and the task) to coordinate effectively~\citep{araujo2016}, non-linear pedagogy encourages players to perceive and exploit affordances -- that is, opportunities for action offered by the environment -- rather than to execute predetermined strategies, thereby promoting the self-organization of an adaptive behavior~\citep{Chow2013, Button2021}.
Based on these two pedagogical conditions, we make the following hypothesis:
\begin{hypothesis_box}{Hypothesis H2}
    The non-linear pedagogy encourages players to explore more various coordination patterns compared to the linear pedagogy. 
\end{hypothesis_box}
Specifically, the exploratory behavior promoted in non-linear pedagogy is expected to result in greater variability in coordination patterns compared to linear pedagogy, both within a given attacking play or between attacking plays of a given set. 

\subsection{Experimental protocol}
The experiment was conducted with 14 male rugby teams, each one composed of 6 players (aged between 15 and 21 years) recruited from 7 elite training centers across France. Each team was randomly assigned to one of the two pedagogical conditions, resulting in 7 teams per pedagogy group. 
The experimental protocol was conducted in three phases. Each team performed 12 attacking plays under the assigned pedagogical principles (i.e. 4 times for each of the 3 defensive scenarios), referred to as the \emph{intervention} phase. This phase was preceded by a \emph{pre} phase, during which teams performed the same 12 attacking plays without any coach intervention, aiming at controlling for any differences between teams \emph{a priori}. Finally, after the intervention, teams entered the \emph{post} phase, during which they performed 20 attacking plays, again without coach intervention, aiming at evaluating how teams behave \emph{a posteriori}, after having been exposed to a specific pedagogical condition. 
Among these 20 attacking plays, teams were confronted 4 times to each of 3 three defensive scenarios already encountered (\emph{post test}) and 4 times to 2 new defensive scenarios, constituting a \emph{transfer test} aiming at revealing how the pedagogical condition in which players were trained impacted their behavior when confronted to new situations. Specifically, the transfer test involved a \emph{specific} transfer (a variation of the kick scenario) and a \emph{general} transfer (a completely new defensive configuration), grounded in the concept of specificity-generality of skill transfer~\citep{seifert2016}. We summarize in Table~\ref{tab:data_set} the data set of the experimental protocol. 

Readers interested to the detailed protocol can refer to~\cite{bourgeais2025}. The protocol design followed the declaration of Helsinki and was approved by the national ethics committee (IRB approval ID:IRB00012476-2022-15-03-163). All the players or parents in case of minors provided their written informed consent. 

\begin{table}[ht]
\centering
\begin{tabular}{c|c|c|c|c|c|c|c|}
\multicolumn{2}{c|}{} & \multicolumn{2}{c|}{\textbf{Pre}} &  \multicolumn{2}{c|}{\textbf{Interv.}} &  \multicolumn{2}{c|}{\textbf{Post}} \\ \cline{3-8}
\multicolumn{2}{c|}{} & L & NL & L & NL & L & NL\\ \hline
\multirow{3}{*}{Training scenarios} & Tight & 20 & 19 & 20 & 20 & 21 & 20 \\ \cline{2-8}
 & Open & 20 & 20 & 20 & 20 & 20 & 18 \\ \cline{2-8}
 & Kick & 19 & 20 & 20 & 20 & 20 & 19 \\ \hline
\multirow{2}{*}{New scenarios} & General & - & - & - & - & 20 & 19 \\ \cline{2-8}
 & Specific & - & - & - & - & 19 & 18 \\ \hline
\end{tabular}
\caption{Number of attacking plays in the data set for each defensive scenario and each pedagogy (Linear: L, Non-linear: NL) during the three phases of the protocol (Pre, Intervention, Post).} 
\label{tab:data_set}
\end{table}

In what follows, we use our model to validate or refute hypotheses \textbf{H1.A}, \textbf{H1.B} and \textbf{H1.C}, and \textbf{H2}, regarding the defensive scenario, respectively the pedagogy.

\subsection{Features}
In this subsection, we define features which enable to answer the validity question of all hypotheses.
First, we focus on hypotheses \textbf{H1.A}, \textbf{H1.B} and \textbf{H1.C}.

\paragraph{Maximum shift right.} 
We define the maximum shift right as a feature measuring the right wing shift of the ball, useful for validating or not hypothesis \textbf{H1.A}, which is as reminder:
\begin{hypothesis_box}{Hypothesis H1.A}
    The tight scenario invites the attackers to pass the ball laterally to bypass the defenders and reach the opposite wing.
    \end{hypothesis_box}
In order to capture the effect of the defensive scenario on this indicator, we measure it only at the \emph{beginning} of the attacking play, let us say for the $k$ first vertices, with $k$ \emph{small}. 

Precisely, we define the maximum shift right reached in the $k$ first vertices of an attacking play as
\begin{equation*}
    \max_{1\leq i \leq k} 5 - r_i\,,
\end{equation*}
where $r_i$ is the value of the second attribute of spatial relative information of vertex $i$, namely is the number of players to the right of the ball carrier at vertex $i$. A large maximum shift right indicates a notable lateral movement toward the right wing of the attacking team. Thus, \textbf{H1.A} amounts to detect large maximum shift right values for the tight defense.

\paragraph{Crossing rank.}
We define a new feature, the crossing rank, as the rank of the first vertex of the attacking play with absolute position ``Middle''. In other words, we quantify the tendency to move across the defense team quickly, in order to validate or not hypothesis \textbf{H1.B}. We recall that the latter hypothesis is:
\begin{hypothesis_box}{Hypothesis H1.B} 
    The open scenario invites the attackers to carry the ball to cross the gaps and break through the defense.
    \end{hypothesis_box}
As for the previous feature, the maximum shift right, we consider only the beginning of the attacking play, namely the first $k$ vertices. The smaller the crossing rank, the quicker the attacking team crossed the gaps and break through the defense. Thus, \textbf{H1.B} amounts to detect small crossing rank for the open defense.

\paragraph{Kick passes.}
Eventually, the last feature we consider, related to hypothesis \textbf{H1.C}, is the number of kick passes. Indeed, this hypothesis is, as a reminder:
    \begin{hypothesis_box}{Hypothesis H1.C} The kick scenario invites attackers to use kick passes to pass over the defense and exploit the space behind them.
    \end{hypothesis_box}
For the $k$ first vertices, we simply compute the number of kick passes. \textbf{H1.C} amounts to observe a larger number of kick passes for the kick defense than for the tight and open defenses.

Next, we focus on hypothesis \textbf{H2} which is:
\begin{hypothesis_box}{Hypothesis H2}
    The non-linear pedagogy encourages players to explore more various coordination patterns compared to the linear pedagogy. 
\end{hypothesis_box}
Validating hypothesis \textbf{H2} amounts to detect a greater variability on coordination patterns under non-linear pedagogy than linear pedagogy. For that, we observe the evolution of the variability during the different phases (pre, intervention and post). Specifically, we measure this variability with two features defined below.

\paragraph{Path length.}
We first consider a \emph{local} feature, i.e. at the scale of an attacking play.
For a given path, representing an attacking play, we define its length as the number of arcs it contains. Path length indicates indistinctly the frequency of reorganisation around the ball carrier (change of spatial relative position), the ball carrier's movement on the field (change of spatial absolute position) and the number of passes (thematic event). The average path length for each pedagogy provides information about the three indications above-mentioned, and can be a discriminant factor.

\paragraph{Subgraph density.}
Contrary to the previous feature, we consider here a \emph{global} feature, i.e. at the scale of a set of attacking plays. For a given set of attacking plays, namely paths on the skeleton graph, we define the density of this subset as the number of arcs of the subgraph representing the unweighted union of all paths divided by the number of all possible arcs in the skeleton graph, i.e. $|V_\text{sp}|^2 + |V_\text{sp}|\cdot|V_\text{res}|$. Note that this number is equal to 360 for the rugby application studied. The density represents the size of the \emph{trace} of the set of paths in the skeleton graph. Thus, it measures variability in the sense that the more distinct the paths are, either totally or partially, two by two, the larger the density.

\subsection{Results on defense constraint}
\label{subsec:defense_constraints}
\begin{figure}[h!]
    \centering
    \subfigure[Distribution of maximum shift right for the first 5 vertices.]{\includegraphics[width=0.4\textwidth]{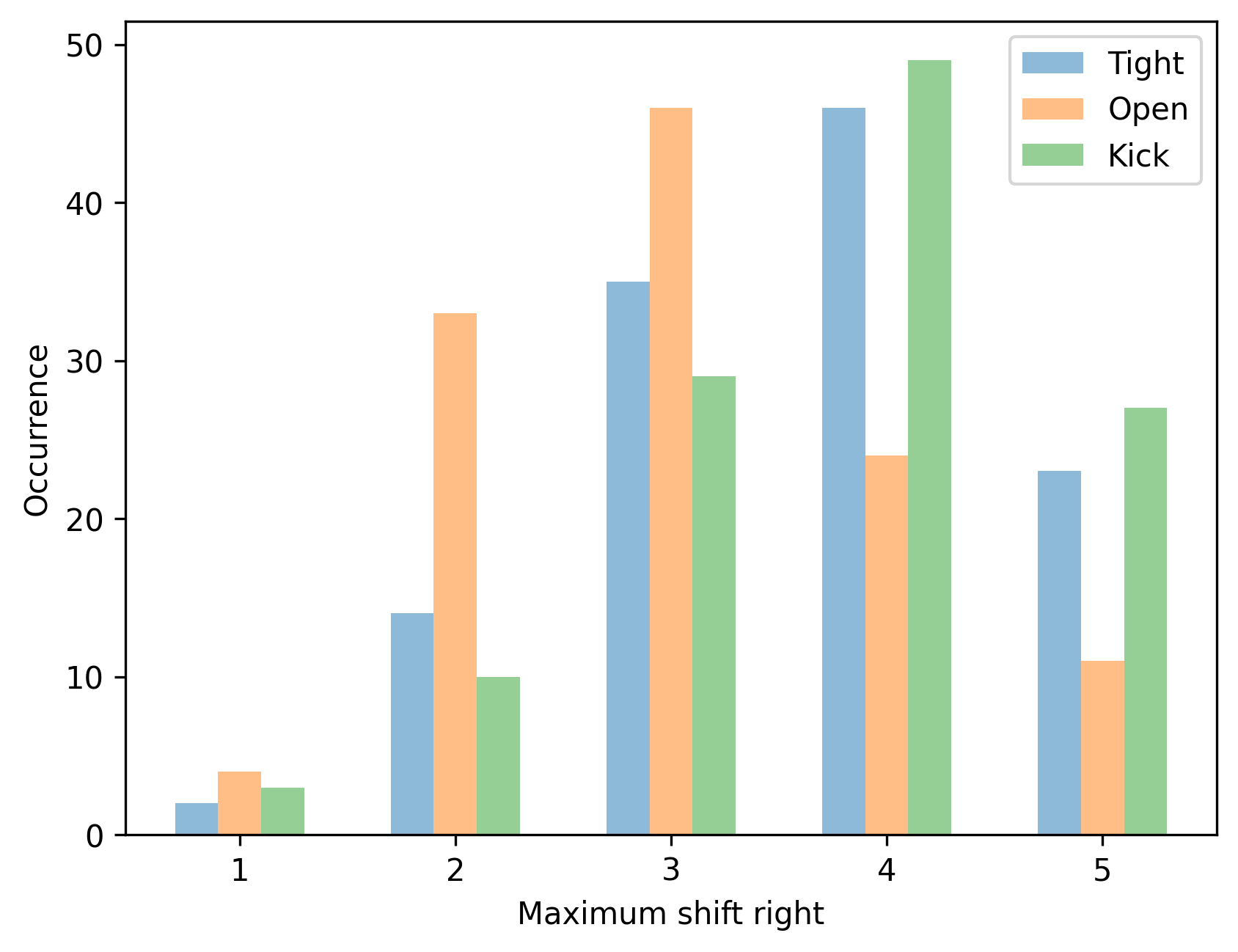}\label{fig:defenses_maximum_shift_right}}
    \hspace{8mm}
    \subfigure[Distribution of crossing rank for the first 5 vertices.]{\includegraphics[width=0.41\textwidth]{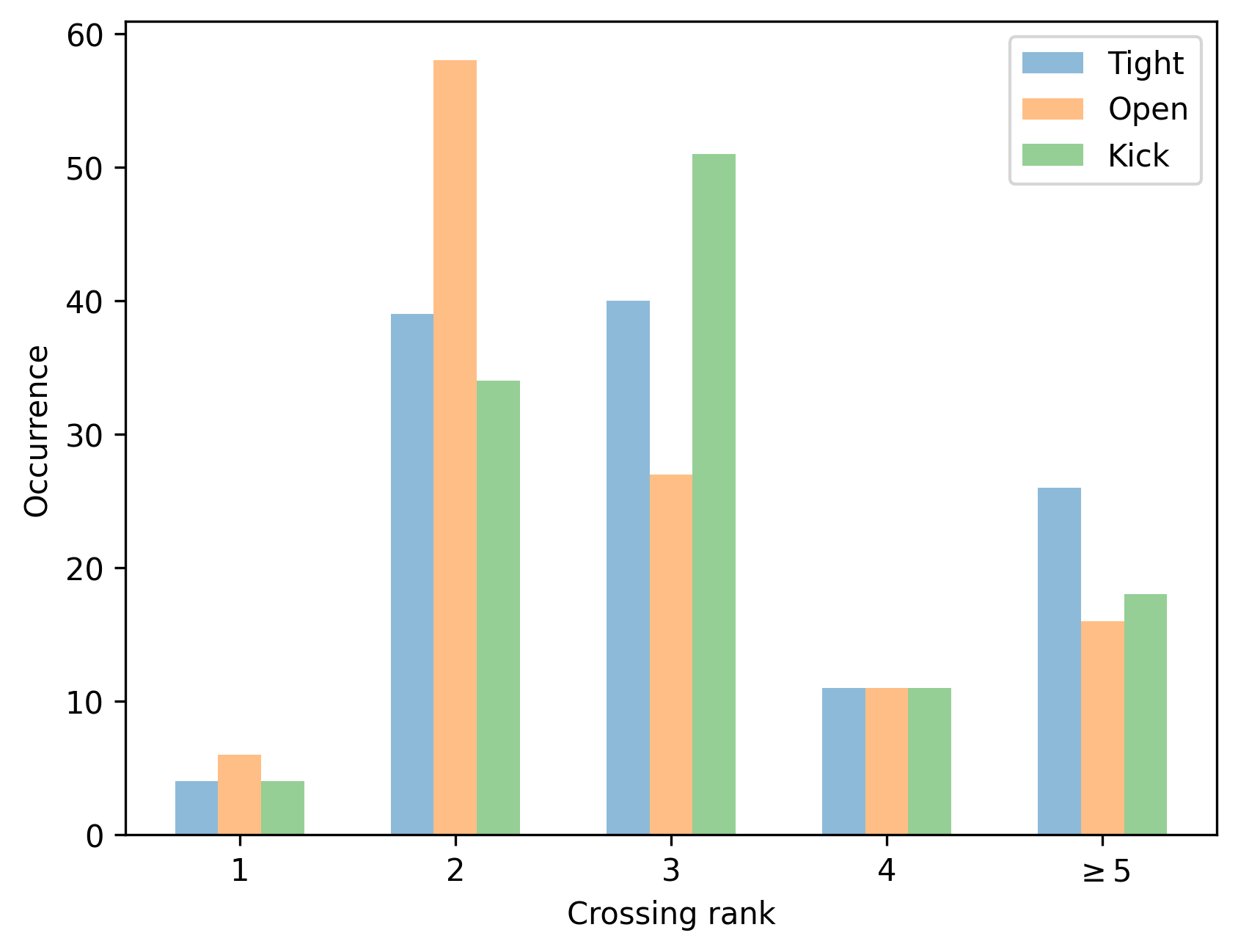}\label{fig:defenses_crossing_speed}} \\
    \subfigure[Number of kick passes for the first 5 vertices.]{\includegraphics[width=0.41\textwidth]{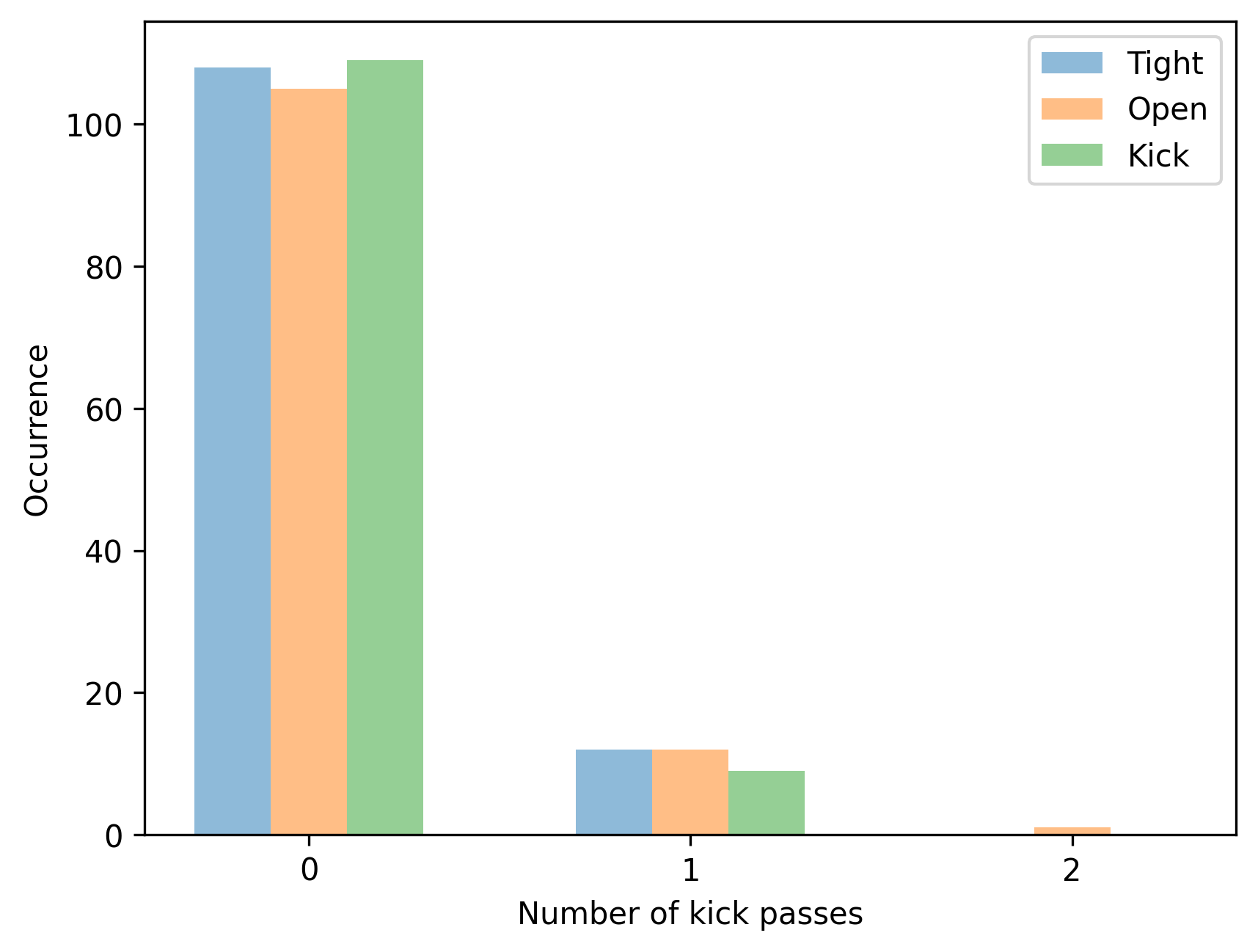}\label{fig:defenses_nb_foot_passes}}
    \caption{Characterization of the impact of the defense over the attacking team behavior over all attacking plays.}
    \label{fig:defenses_all_features}
\end{figure}

First, we report the distribution of maximum shift right, crossing rank, and kick passes in Figure~\ref{fig:defenses_all_features}, for each of the three defensive scenarios (tight, open, kick), for the first $k=5$ vertices of attacking plays, considering indistinctly all attacking plays, whatever the pedagogy or the phase it relates to. Notice that the number of $k=5$ vertices, i.e. working on the sub-path starting at the initial vertex and of length 4, has been chosen to represent half of the average path length (equal to 8 for pre phase). Indeed, we suppose that the initial position of the defense team can no longer have a meaningful effect after half of the average path length of observed attacking plays. By definition, maximum shift right is indexed on the number of players and values therefore range from 1 to 5. For crossing rank, values are not intrinsically limited, but they range from 1 to 5 here since we are limiting to $k = 5$. The attacking plays labeled as ``$\geq 5$'' refer to instances where the absolute position ``Middle'' is not achieved within the first 5 vertices considered. This occurs either because the position is reached later in the path or because the defense has not been broken. Finally, the number of kick passes is predominantly zero. Overall, the event is very rare, and when it does occur, it happens once, or twice in a very few cases.

To examine the effect of defensive scenarios on team coordination patterns, observations of the respective distributions of the three features are associated with non-parametric statistical tests. For maximum shift right and crossing rank, differences are evaluated using Kruskal–Wallis tests, with effect sizes reported using epsilon squared ($\epsilon^2$) and their 95\% confidence intervals. In the case of a significant main effect, pairwise comparisons were performed using Dunn’s tests with Holm correction for multiple comparisons, using rank-biserial correlations ($r_\text{rb}$) based on individual Mann-Whitney tests as measures of effect size. For the number of kick passes, it is treated as a binary variable (i.e. presence vs. absence of at least one kick pass within the first five vertices) and the association with defensive scenario is examined using a chi-square independence test. All statistical analyses were conducted using JASP (software version 0.19.3.0), and the significance threshold was set at $\alpha = 0.05$.

\paragraph{Maximum shift right.}
The results of Figure~\ref{fig:defenses_maximum_shift_right} show that the open defense scenario leads to a smaller maximum shift right than the tight and kick defense scenarios, meaning that these two latter invite to move quickly in width to the right. Indeed, computing the median value of all possessions of the three scenarios indicates that half of the attacking plays have a maximum shift right smaller or equal to 3, and thus half of the attacking plays correspond to a maximum shift right larger or equal to 4. Moreover, the percentage of attacking plays with \emph{large} maximum shift right values (i.e. 4 or 5 according to the median) for each defense is: 56\% for tight scenario; 30\% for open scenario; 64\% for kick scenario. Thus, we observe that large values over-represent the tight and kick scenarios, on the contrary to the open scenario which is under-represented. 

This is supported by the Kruskal–Wallis test that reveals a significant effect of defensive scenario on maximum shift right ($\chi^2(2) = 32.75, p < .001$) with a moderate effect size ($\epsilon^2 = 0.092$, 95\% CI [0.045, 0.157]). Specifically, the open scenario yields significantly lower maximum shift right values than both the kick scenario ($z = -5.38$, $p < .001$, $r_\text{rb} = 0.385$) and the tight scenario ($z = -4.39$, $p < .001$, $r_\text{rb} = 0.320$), while there is no significant difference between the kick and tight scenarios ($z = 1.01$, $p = .315$, $r_\text{rb} = 0.076$).

\paragraph{Crossing rank.}
We observe in Figure~\ref{fig:defenses_crossing_speed} that around half of all the attacking plays have a crossing rank smaller or equal to 2. In other word, a median split would occur between the values 2 and 3. Yet, 54\% of the attacking plays for the open scenario have a value equal to 1 or 2 (i.e. \emph{small} values of crossing rank according to the median), whereas this percentage goes down to 36\% and 32\% for the tight, respectively kick, scenario. Thus, we detect a difference between the open scenario and the tight and kick scenarios: the open scenario over-represents small crossing rank values on the contrary to tight and kick scenarios which are represented to larger values in average. 

This is again supported by the Kruskal–Wallis test, revealing a significant effect of defensive scenario on crossing rank ($\chi^2(2) = 8.65$, $p = .013$), but only with a small effect size ($\epsilon^2 = 0.024$, 95\% CI [0.002, 0.066]). Specifically, the open scenario yields significantly lower crossing rank values than both the kick scenario ($z = -2.41$, $p = .032$, $r_\text{rb} = 0.177$) and the tight scenario ($z = -2.67$, $p = .023$, $r_\text{rb} = 0.186$), while there is no significant difference between the kick and tight scenarios ($z = -0.25$, $p = .800$, $r_\text{rb} = 0.023$).

\paragraph{Kick passes.}
The results of Figure~\ref{fig:defenses_nb_foot_passes} do not show any difference between the three defensive scenarios in term of number of kick passes. This conclusion is not surprising because, according to the rugby experts consulted during the study and after analyzing the protocol experiments data, kick passes are rare events, hard to execute, and thus the number of kick passes is too small (34 kick passes over 356 attacking plays) to be reliable on expressing the intention of the attacking team. 

The chi-square test confirmed the absence of significant association between defensive scenario and the presence of at least one kick pass within the first five vertices ($\chi^2(2, N = 356) = 0.07$, $p = .968$).

\subsection{Results on pedagogy constraint} 
\label{subsec:pedagogy_constraints}
\begin{figure}[h!]
    \centering
    \includegraphics[width=0.8\linewidth]{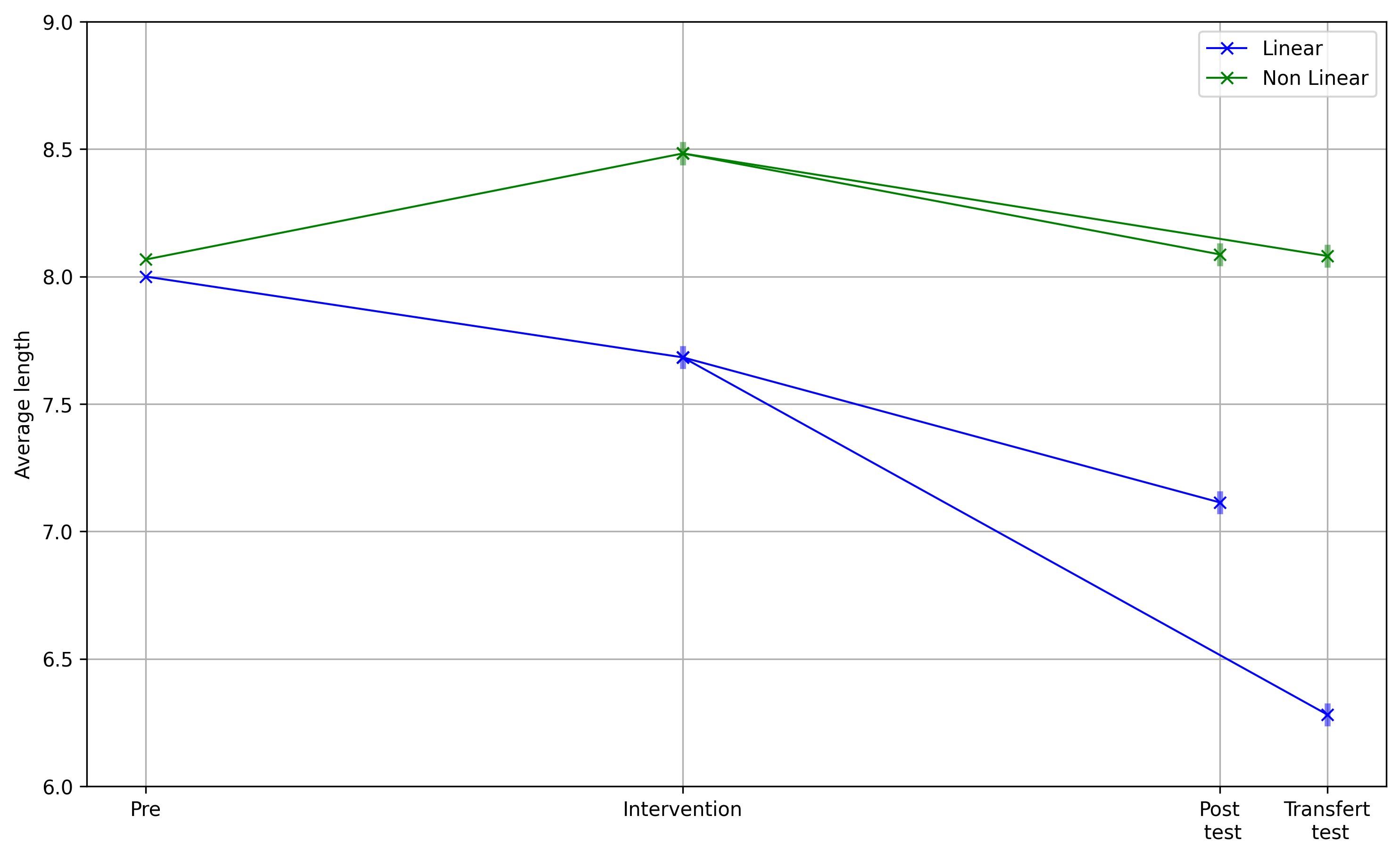}
    \caption{Evolution of the average path length for each pedagogy. The error bar represents randomness in protocol (based on the difference observed between groups in pre phase).}
    \label{fig:pedagogies_average_length}
\end{figure}

\begin{figure}[h!]
    \centering
    \includegraphics[width=0.8\linewidth]{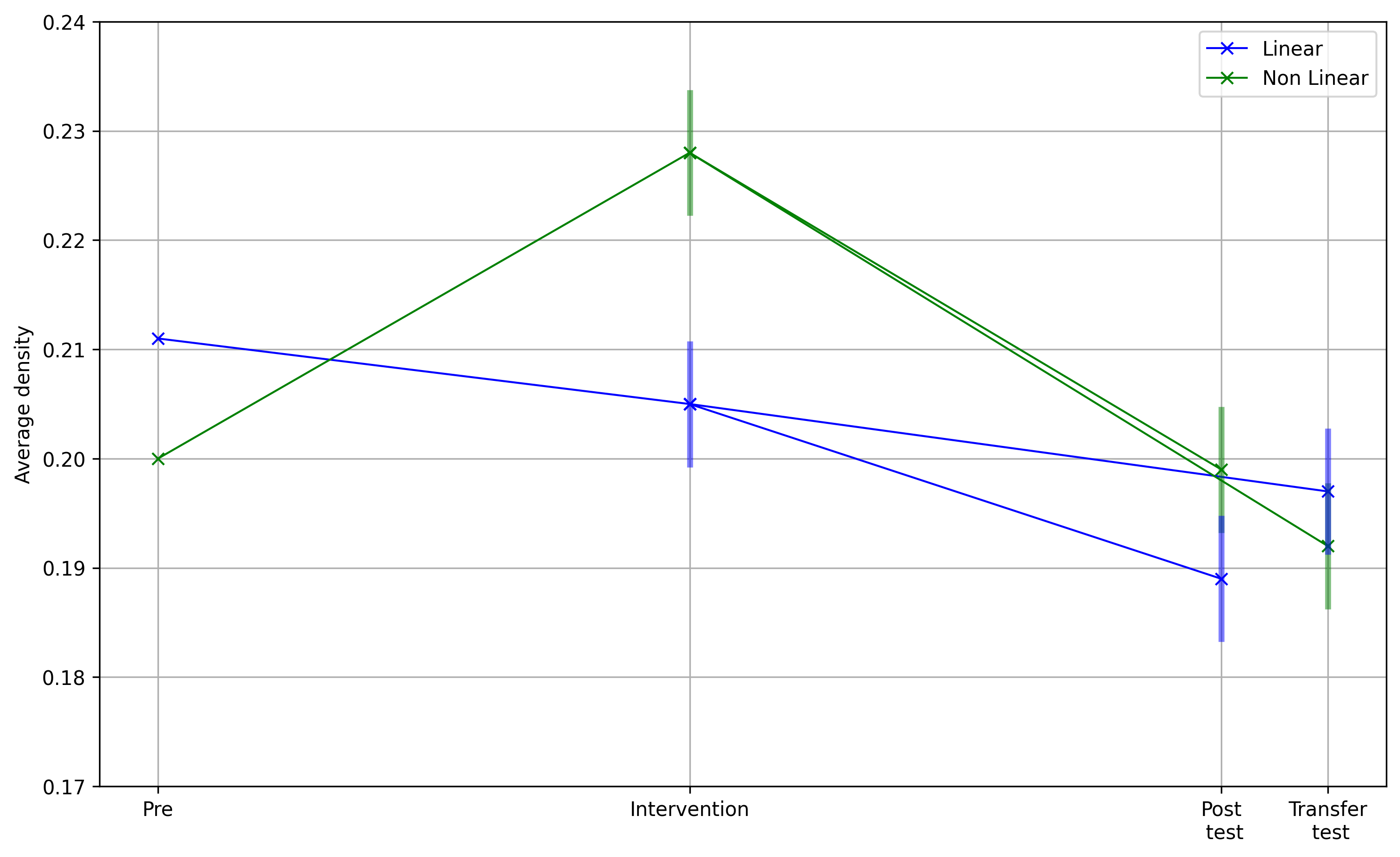}
    \caption{Evolution of the average subgraph density, over the different defenses, for each pedagogy. The error bar represents randomness in protocol (based on the difference observed between groups in pre phase).}
    \label{fig:pedagogies_average_density}
\end{figure}

To evaluate the effect of the pedagogical condition on team coordination patterns, we display the evolution of the average path length (Figure~\ref{fig:pedagogies_average_length}) and of the average subgraph density (Figure~\ref{fig:pedagogies_average_density}) for each pedagogy and at each phase (i.e. the defensive scenarios are not distinguished). Precisely, we compute these features for the set of all attacking plays of a given pedagogy and a given phase. Notice that for the post phase, we distinguish the attacking plays between post test and transfer test.

By definition, both metrics are computed based on aggregated data and do not allow statistical tests to be performed to support the observations. Hence, for both features, we use the difference between linear and non-linear pedagogies at the pre phase to provide confidence intervals. Thus, it characterizes the randomness in protocol, and is represented by an error bar in the following phases (intervention and post).

\paragraph{Path length.} We observe in Figure~\ref{fig:pedagogies_average_length} that the difference of the average path length of the two groups observed during the pre phase increases during the intervention phase (the difference is 12 times larger during the intervention phase), indicating a notable effect of the pedagogy over this feature. Indeed, the average path length for non-linear pedagogy increases by $+4.9\%$ whereas the average path length for the linear pedagogy decreases by $-4.1\%$. Precisely, the average path length is lower for the linear pedagogy group than for the non-linear one. This indicates that the variability of spatial and/or thematic information increases when the coach provides non-linear pedagogy conditions. This difference remains in the post phase, namely when the players are left without any coach intervention, and we can note that only the linear pedagogy group presents a difference between the defensive scenarios. Indeed, the average path length is larger (of $11.7\%$) for already encountered scenarios (post test) than for new defenses (transfer test). This shows that, on this feature, being confronted to a new defensive scenario affects the linear pedagogy group contrary to the non-linear one.

\paragraph{Subgraph density.} We observe in Figure~\ref{fig:pedagogies_average_density} that the difference of the subgraph density between the two groups during the pre phase increases by $+52.2\%$ during the intervention phase, indicating that the pedagogy has a notable influence on this feature. Specifically, the subgraph density for the non-linear pedagogy group ($+12.3\%$ compared to the pre phase) is larger than the one of linear pedagogy group ($-2.9\%$ compared to the pre phase), indicating a larger variability on coordination patterns. However, on the contrary to previous feature average length path, we do not observe meaningful difference during post phase. 

\subsection{Discussion about experiments}
\label{subsec:discussion_results}

Firstly, let us discuss the validity of hypotheses \textbf{H1.A}, \textbf{H1.B} and \textbf{H1.C}.
The effects of the defensive scenarios on the coordination patterns partly verify these hypotheses: based on the graph features we have defined, the tight and kick scenarios appear rather similar, while the open scenario is different. Specifically, hypothesis \textbf{H1.A} is partly validated by the analysis of the maximum shift right feature: the tight defense invites attackers to pass the ball laterally to bypass the defenders and reach the right wing, but the same effect is observed with the kick defense. Hypothesis \textbf{H1.B} is validated by the analysis of the crossing rank feature, indicating that players carried more the ball to cross the gaps between defenders and break through the defense in the open defense compared to the others scenarios.  Eventually, hypothesis \textbf{H1.C} could not be verified, since the analysis of the number of kick passes yields no differences, but it may be due to the very low occurrence of this type of event in the set of observed attacking plays.

Secondly, we discuss the validity of hypothesis \textbf{H2}.
The effects of the pedagogical approach on the coordination patterns verify hypothesis \textbf{H2}, as the results indicates that non-linear pedagogical principles lead to greater variability in coordination patterns than linear pedagogy principles. More specifically, this effect is observed at both local and global scales: greater variability is observed in path length and subgraph density under non-linear pedagogy, indicating more diverse coordination patterns both within and between attacking plays. This suggests that teams explore a wider range of solutions and adapt more their organization to the dynamical game situations -- both in terms of ball movement (i.e. how and where the ball moved through passes or carries) and spatial organization of players around the ball (i.e. their relative positioning). These results are therefore aligned with the outcomes expected from the application of non-linear pedagogy principles~\citep{chow2006, lee2014, gray2020, orangi2021}. It illustrates how non-linear pedagogy fosters adaptability and exploration, enabling teams to self-organize and adjust their collective behavior more effectively than under linear pedagogy. In the context of our work, this result may be explained by the fact that non-linear pedagogy encourages players to explore more the game environment and attune to shared affordances that emerges from player—player—environment interactions~\citep{silva2013}, thereby possibly enhancing tactical perception, positioning, and coordinated movement between teammates.
At the inter-possession scale (corresponding to subgraph density feature), the difference in variability between pedagogical approaches disappear in the post phase, in which players were asked to self-organize again, without coach guidance. This indicates the presence of a coach providing collective organization principles to players is still required to enhance such variability. In contrast, the intra-possession variability (corresponding to path length feature) effect is persistent: the capability to reorganize during the attacking play itself remains higher in the non-linear pedagogy group, even without coach guidance. This is true for non-linear pedagogy groups across both training and new scenarios of defense, whereas players trained under linear pedagogy show a clear decrease in variability when facing new scenarios, suggesting that their coordination patterns remain even more rigid beyond the trained situations. These findings thus support the theoretical distinction between the two pedagogical approaches: while linear pedagogy fosters the reproduction of pre-defined solutions, non-linear pedagogy promotes the emergence of adaptive coordination patterns through exploration and self-organization, allowing players to adapt their collective behavior to evolving game situations. Despite this, these results remains limited, mostly due to too short intervention time, thus not allowing for a true skill acquisition and skill transfer. 

However, some methodological limitations should be acknowledged when interpreting the results. First, the ecological nature of the experimental protocol leads to practical constraints that may have affected the quality of the data itself. Contextual factors such as weather conditions, players' specific skills, technical errors, and time restrictions occasionally introduce uncontrolled variability -- particularly in a context with such a limited number of teams and attacking plays. Second, the influence of the coach constitutes a major source of variability: the experimental protocol involved several coaches, each implementing both pedagogical approaches with two different groups of players. Consequently, differences between coaches may have affected the manipulation of pedagogy constraints, as it was dependent on how each coach interpreted, understood, and executed linear and non-linear pedagogical principles. This is all the more true since previous research has shown that there are differences between what coaches think they do and what they actually do in practice~\citep{ashford2025}. These elements underline the inherent challenge of conducting controlled experiments under realistic training conditions with elite players, inevitably leading to a limited set of attacking plays, which must be kept in mind when interpreting the results.

\section{Conclusion and perspectives}
\label{sec:discussion}

In this paper, we propose a new spatio-temporal graph-based model of team sports which constitutes a tool for testing the validity of hypotheses on team coordination patterns. This model thus constitutes a practical tool for team sports analysis and specifically for answering concrete questions about team coordination. It has shown promising results on the various hypotheses tested in this work, even if the tiny size of the data due to the complexity of setting an experiment protocol does not enable a strong assertion but more a general trend validating the hypotheses.
Note that the experimental protocol presented in this paper was not designed to validate the model's effectiveness but to answer some of these questions. In other words, we evaluate whether some hypotheses corresponding to these questions have been supported or refuted by our model (see Subsection~\ref{subsec:discussion_results}) but we have not yet discuss the model itself. Let us then delve into it and specifically the choices made in its design representing different leverages.

First, the choice of the parameters of the model are of great importance: the zoning for absolute and relative spatial positions, and the set of thematic events $\{th_1,\ldots, th_k\}$. On the one hand, the number of absolute and relative zones on the field directly impacts the size of the skeleton graph and consequently the size of the paths. On the other hand, the set of thematic events also influences the multiplicity of paths that the model can output. Such parameters should be chosen according to the characteristics of the team sport in question. Future work should focus on testing other configurations, in terms of zoning and thematic events, to measure the impact of these parameters on the results, and thus to assess the robustness of the model. Indeed, one challenge for this model is to find relevant parameters enabling to answer team coordination questions of very different nature under a parsimonious setting.

Second, the output of the model can be analyzed from different perspectives. As previously discussed in~\ref{subsec:measures}, we can consider outputs at two different scales: local scale, which is a path representing a particular attacking play; and global scale, which is a subgraph of the skeleton graph representing a set of attacking plays. In this paper, we study both local scale with path length and global scale with subgraph density. In addition to the scale we choose to study the output, we also determine how to analyze it. In this paper, we only use features, namely absolute indicators, to characterize the outputs under study and compare them in an absolute manner. However, comparing two outputs (at the same scale) relatively with metrics -- or pseudo-metrics -- is ongoing work for future research on new experimental protocols. Due to the nature of the outputs (absolute and relative information in each vertex, semantic labels on arcs, paths of different sizes, etc.), the definition of such metrics is non-trivial. For instance, a pseudo-metric at local scale that could be considered is the maximal common subpath: given two paths, it determines the common subpath of the largest size. This raises the question of the \emph{equality} or \emph{likeness} between two paths. Then, a relevant distance between two vertices, two arcs, and then two paths can be established. However, one additional challenge is that it can be done in multiple various ways, which represent active research work. It is worth noting that the definition of relevant metrics opens the possibility to discriminate between certain team coordination patterns, using learning methods for instance.

Third, the relevance of our model can be improved with new information. In the proposed model, we decide to consider information relating only to the attacking team because we study only internal team coordination effects. However, it could be relevant to integrate the opposite team as a new source of information, as team sports phenomena intrinsically integrate the relation to the other team~\citep{Buldu2018}. A first lead to do so would be to extend the definition of vertices of the skeleton graph to integrate the relative position of opposite team players -- certainly with a different partitioning of the field. 

Eventually, undergoing study on basketball with slightly different experimental protocol, from which the data observed are being currently extracted and treated to be modelized by the model, will enable a deeper comprehension of the three main points above-mentioned. Indeed, basketball and rugby are two team sports based on radically different intern logic: rugby is driven by the progression to the field toward the try line whereas basketball is driven by managing to make free space between the ball carrier and the basket in order to shoot. These fundamental differences will lead to consider different zoning and thematic events (e.g. absolute zoning related to key, 2-point or 3-point zones; relative zoning naturally oriented toward the basket etc.) together with different metrics at different scales.

\section{Declarations}
\paragraph{Conflict of interest.}
We declare no conflict of interests.

\paragraph{Fundings.} 
We acknowledge the support of the French National Agency of Research (grant ID: ANR-19-STHP-0006 TEAM SPORTS and ANR-23-CE38-0008 DYNATEAM).

\bibliographystyle{apalike}
\bibliography{biblio}

\end{document}